\documentclass[]{aa} 

\usepackage[switch]{lineno}

\usepackage{graphicx}
\usepackage{txfonts}
\usepackage[]{hyperref}
 \usepackage{graphicx}
 \usepackage[normalem]{ulem}
 \useunder{\uline}{\ul}{}
 \usepackage{lscape}
 \usepackage{xcolor}
 \usepackage{float} 
 \usepackage[colorinlistoftodos]{todonotes}

\begin{document}

    \title{The multi-spacecraft high-energy solar particle event of 28 October 2021}

    \author{
        A. Kouloumvakos \inst{1}
        \and
        A. Papaioannou \inst{2}
        \and
        C. O. G. Waterfall \inst{3}
        \and
        S. Dalla \inst{3}
        \and
        R. Vainio \inst{4}
        \and
        G. M. Mason \inst{1}
        \and
        B. Heber \inst{5}
        \and
        P. K\"{u}hl \inst{5}
        \and
        R. C. Allen \inst{1}
        \and
        C.M.S. Cohen \inst{6}
        \and
        G. Ho \inst{1}
        \and
        A. Anastasiadis \inst{2}
        \and
        A. P. Rouillard \inst{7}
        \and
        J. Rodríguez-Pacheco \inst{8}
        \and
        J. Guo \inst{9,10}
        \and
        X. Li \inst{9}
        \and
        M. H\"{o}rl\"{o}ck \inst{5}
        \and
        R. F. Wimmer-Schweingruber \inst{5}
        }
       
    \institute{
        The Johns Hopkins University Applied Physics Laboratory, 11101 Johns Hopkins Road, Laurel, MD 20723, USA \\
        \email{Athanasios.Kouloumvakos@jhuapl.edu}
        \and
        Institute for Astronomy, Astrophysics, Space Applications and Remote Sensing (IAASARS), National Observatory of Athens, I. Metaxa \& Vas. Pavlou St., 15236 Penteli, Greece
        \and
        Jeremiah Horrocks Institute, University of Central Lancashire, Preston, PR1 2HE, UK
        \and
        Department of Physics and Astronomy, University of Turku, 20500 Turku, Finland
        \and
        Institut f\"ur Experimentelle und Angewandte Physik, Christian-Albrechts-Universit\"at zu Kiel, 24118 Kiel, Germany
        \and
        California Institute of Technology, Pasadena, CA, USA
        \and
        IRAP, CNRS, Université Toulouse III–Paul Sabatier, Toulouse, France
        \and
        Universidad de Alcalá, Space Research Group, 28805 Alcalá de Henares, Spain
        \and
        Deep space Exploration Laboratory/School of Earth and Space Sciences, University of Science and Technology of China, Hefei, Anhui, 230026, PR China
        \and
        CAS Center for Excellence in Comparative Planetology, USTC, Hefei, PR China
                }

\titlerunning{The 28 October 2021 SEP event}  

 
\abstract
   {}
   {We studied the first multi-spacecraft high-energy solar energetic particle (SEP) event of solar cycle 25, which triggered a ground level enhancement (GLE) on 28 October 2021, using data from multiple observers (Parker Solar Probe, STEREO-A, Solar Orbiter, GOES, SOHO, BepiColombo, and the Mars Science Laboratory) that were widely distributed throughout the heliosphere and located at heliocentric distances ranging from 0.60 to 1.60 AU.}
   {We present SEP observations at a broad energy range spanning from $\sim$10 to 600 MeV obtained from the different instruments. We performed detail modelling of the shock wave and we derived the 3D distribution and temporal evolution of the shock parameters. We further investigated the magnetic connectivity of each observer to the solar surface and examined the shock's magnetic connection. We performed velocity dispersion analysis (VDA) and time-shifting analysis (TSA) to infer the SEP release time. We derived and present the peak proton flux spectra for all the above spacecraft and fluence spectra for major species recorded on board Solar Orbiter from the Suprathermal Ion Spectrograph (SIS). We performed 3D SEP propagation simulations to investigate the role of particle transport in the distribution of SEPs to distant magnetically connected observers.}
   {Observations and modelling show that a strong shock wave formed promptly in the low corona. At the SEP release time windows, we find a connection with the shock for all the observers. PSP, STEREO-A, and Solar Orbiter were connected to strong shock regions with high Mach numbers (>4), whereas the Earth and other observers were connected to lower Mach numbers. The SEP spectral properties near Earth demonstrate two  power laws, with a harder (softer) spectrum in the low-energy (high-energy) range. Composition observations from SIS (and near-Earth instruments) show no serious enhancement of flare-accelerated material.}
   {A possible scenario consistent with the observations and our analysis indicates that high-energy SEPs at PSP, STEREO-A, and Solar Orbiter were dominated by particle acceleration and injection by the shock, whereas high-energy SEPs that reached near-Earth space were associated with a weaker shock;   it is likely that efficient transport of particles from a wide injection source contributed to the observed high-energy SEPs. Our study cannot exclude a contribution from a flare-related process; however, composition observations show no evidence of an impulsive composition of suprathermals during the event, suggestive of a non-dominant flare-related process.}


   \maketitle
%

\section{Introduction} \label{sec:intro}

Acceleration of high-energy particles at the Sun is a challenging issue in solar and space physics research, and there is a long-standing debate about the mechanisms that can accelerate particles to energies ranging from a few tens of keVs to several GeVs a few minutes after the start of the eruption. Solar energetic particle (SEP) events are typically associated with solar jets, flares, coronal mass ejections (CMEs), and shock waves, and are a key ingredient of solar and heliospheric physics research. Several different physical mechanisms can be responsible for the energization and acceleration of SEPs  \citep[see e.g.][for recent reviews]{Anastasiadis2019, Vlahos2019, Klein2019, Reames2021} during these events. The three main processes that could lead to an efficient acceleration of particles \citep{Petrosian2008, Vainio2018} are a 1) direct acceleration by electric fields associated with reconnection or induction by large-scale magnetic fields, 2) stochastic acceleration (second-order Fermi acceleration) in turbulence or by plasma waves, and 3) {diffusive shock (first-order Fermi acceleration) or compressional acceleration}.

High-energy SEP events are associated with intense flares and with fast, wide, and strong CME-driven shock waves \citep{Rouillard2012, Rouillard2016, Kouloumvakos2019}. Many studies argue that both flare-related and shock-related acceleration processes can contribute to high-energy SEP events \cite[e.g.][]{Cane2006, Cane2007, Kouloumvakos2015, Papaioannou2016, SalasMatamoros2016, Zhao2018, Kocharov2021}, whereas  several others argue that one of the two processes (magnetic reconnection or shocks) probably dominates in some high-energy SEP events \citep[e.g.][]{Klein2001, Simnett2006, Klein2014, Kouloumvakos2020}. It has been suggested that magnetic reconnection at the current sheet underneath the CME and/or at places where the CME interacts with the ambient coronal magnetic field can accelerate protons to very high energies \citep{Klein2001, Klein2014} on a very short timescale. These particles may escape onto open magnetic field lines {when the magnetic field of the CME reconnects with the ambient coronal magnetic field} \citep{Masson2013}. There are   other studies that indicate that CME-driven shock waves could have an important role in accelerating SEPs to high energies \citep[e.g.][]{Reames2013, Rouillard2016, Plotnikov2017, Kouloumvakos2019}. Self-consistent SEP modelling of diffusive shock acceleration, which is considered to be the main mechanism in shock acceleration \citep{Bell1978, Blandford1978}, has shown that CME-driven shocks can accelerate SEPs from a few hundred keV to several GeV \citep[][]{Afanasiev2018} in a few minutes. Additionally, turbulence that develops in large-scale coronal loops during the global magnetic field reconfiguration phase or at the shock sheath region is also a possible mechanism that can accelerate protons to high energies.

Observational studies show that many widespread SEP events are associated with fast and wide shock waves that {are capable of accelerating and injecting particles} over a broad range of longitudes \citep{Rouillard2012, Lario2014, Zhu2018, Kouloumvakos2022}. Several other studies show that ground-level enhancement (GLE) events in which ions are accelerated to relativistic energies are associated with fast shocks in the solar corona \citep{Reames2009, Gopalswamy2013, Zhu2021, Liu2019} and suggest that GLE SEPs are accelerated predominately in CME-driven shocks \citep[][]{Kahler2012, Nitta2012, Kouloumvakos2020}. These shocks can be supercritical and strong for a long period after the start of the eruption and over a wide extent \citep[see e.g.][]{Kwon2017, Kouloumvakos2020}. The interaction of shocks with streamers probably favours particle trapping, and hence increases the shock acceleration efficiency \citep{Kong2017,Kong2019}. This shock interaction with coronal structures and mostly streamers seems to play an important role in the acceleration of high-energy SEPs \citep{Morosan2019, Frassati2022}. Advanced shock reconstruction and modelling techniques \citep{Kwon2014, Rouillard2016, Jin2018, Plotnikov2017, Kouloumvakos2019} that provide shock parameters along the 3D shock surface suggest that the shock strength is an important parameter that  characterizes the particle acceleration efficiency of the shock waves \citep[see][for an association with the low energies]{Pacheco1998}. The association of strong shocks with large, high-energy, SEP events was highlighted by the strong correlation found by \cite{Kouloumvakos2019} between the  20 and 100~MeV proton peak intensities and the shock Mach number at magnetically well-connected regions to the observing spacecraft that was determined from the modelling of the associated shock waves. Similar results have been reported for high-energy electrons \citep{Dresing2022}. Lastly, a recent study of the most {longitudinally} distant behind-the-limb flare ever detected in >100~MeV gamma rays by Fermi-LAT \citep{Pesce-Rollins2022} showed that the onset of a coronal shock wave on the visible disk was in coincidence with the LAT onset, which is an unambiguous detection of high-energy particles accelerated by a shock wave.

In addition to the properties of the acceleration, particle transport in the interplanetary medium also plays a role in determining SEP spatial distributions and observables in the heliosphere, such as the time-intensity profiles. {Interplanetary CMEs and stream interaction regions are among the structures that can modify the interplanetary magnetic field and change the SEP transport conditions and that can modify the SEP intensity--time profiles \citep[e.g.][]{Wijsen2020, Wijsen2023}. Turbulence in the interplanetary space is another important element that} produces scattering, often described via a diffusive approach {\citep[e.g.][]{Zhang2009, Droge2010}}, as well as magnetic field line meandering \citep[e.g.][]{Laitinen2016}. Drifts associated with  the gradient and curvature of the average interplanetary magnetic field \citep{Dalla2013, Marsh2013} {can produce transport across the magnetic field. In addition, heliospheric current sheet drift 
\citep{Battarbee2018,Waterfall2022} may give rise to significant particle displacement in longitude and latitude (depending on heliospheric current sheet (HCS) inclination) away from the injection location.} These processes are energy dependent, so that their relative contribution to  propagation depends on the energy of the SEPs under study. {For example, gradient and curvature drift effects are more prominent at high energies \citep{Dalla2013}.}

A remaining open issue in SEP studies is to quantify the contribution, if any,  of each acceleration process to each particle species and to a broad energy range, and to determine if one mechanism systematically dominates over the others and under which particular conditions. {Every scenario} has weaknesses and has been criticized based on various observations. For example, the scenario that a flare-related process dominates implies that distinct characteristics of impulsive SEPs such as enhancements of $^3$He or high Fe/O ratios should be observed in most of the major high-energy SEP events, which is {not the case \citep[see][]{Kahler2012}.} The escape of the SEPs from the flaring region where the closed magnetic topology dominates {is another issue \citep[see][]{Reames2013}, with  modelling studies that suggest that flare-accelerated particles trapped in the CME can gain access to open field lines by reconnections between the CME's flux rope and the ambient field \citep{Masson2013}} On the other hand,   shock acceleration is criticized mainly because of the observed discrepancies with the timings of the SEP release times and the evolution of the shock and the anisotropy {characteristics of the SEPs that do not always agree with the expectations \citep{Miteva2014}}. In both cases, particle diffusion or particle trapping may need to be assumed for some events to explain the observations. Nevertheless, {for some events it is difficult to interpret multi-spacecraft observations with only one acceleration region and mechanism \citep{SalasMatamoros2016}, which means that more that one mechanisms can apply \citep{Papaioannou2016}}.

In this paper we study the high-energy solar particle event of 28 October 2021, which is the first {multi-spacecraft GLE (GLE73) event} of solar cycle 25. The GLE event at Earth and near-Earth measurements of the event were reported by \citet{Papaioannou2022}. Moreover, \cite{Klein2022} performed  radio observations during the event and examined the role of the expanding CME to the GLE acceleration and release. In addition, \cite{Mishev2022} discussed the differences between deka-MeV and high-energy protons. In our study we take advantage of multi-spacecraft data from Solar Orbiter \citep[SolO;][]{Muller2020}, Parker Solar Probe \citep[PSP;][]{Fox2016}, BepiColombo \citep{Benkhoff2010}, and the Mars Science Laboratory \citep[MSL;][]{Hassler2012}. We use advanced shock reconstruction and modelling techniques to determine the shock properties during the event and examine the role of the CME-driven shock wave to the acceleration of {high-energy (GLE-level)} SEPs. Simulations with a 3D test particle code are used to investigate transport effects. Our aim is to {gain further} insight into where the high-energy SEPs accelerated at the Sun, when they were released from their sources to interplanetary space and how they were transported into the heliosphere, and what is the role of the shock wave into these processes during the first multi-spacecraft high-energy event of solar cycle 25. 


\section{Instrumentation}

For this multi-spacecraft study, we analysed observations from instrumentation on board different spacecraft. We used data from instruments on board Solar Orbiter \citep[SolO;][]{Muller2020}, Parker Solar Probe \citep[PSP;][]{Fox2016}, Solar TErrestrial RElations Observatory \citep[STEREO;][]{Kaiser2008}-A, SOlar and Heliospheric Observatory \citep[SOHO;][]{Domingo1995}, Solar Dynamics Observatory \citep[SDO;][]{Pesnell2012}, and the Geostationary Operational Environmental Satellite (GOES).

We provide a brief summary of the data used in this study describing first the measurements of two new solar missions. From SolO, we employed measurements of energetic particles from the Energetic Particle Detector \citep[EPD;][]{Pacheco2020,Wimmer2021} instrument suite, which contains multiple sensors. In this study we used data of energetic protons from the High Energy Telescope (HET) at an energy range from $\sim$10 MeV/nucleon to above $\sim$100 MeV/nucleon. These particle recordings from HET are used primarily for the determination of the SEP onset times. Additionally, we used SEP composition observations from the Suprathermal Ion Spectrograph (SIS) of SolO. From PSP we used particle observations provided by the Integrated Science Investigation of the Sun \citep[IS$\sun$IS;][]{McComas2016} instrument suite. We utilized energetic particle measurements from the Energetic Particle Instruments (EPI). The two IS$\sun$IS/EPI measure the lower (EPI-Lo) and higher (EPI-Hi) energy parts of the energetic particle distributions. In this study we focus on the high-energy part, and we use data from the Low Energy Telescope (LET) and the High Energy Telescope (HET) of EPI-Hi that measures ions from $\sim$1--200 MeV/nucleon. We also used observations from STEREO-A HET \citep[][]{Rosenvinge2008}, BepiColombo Environment Radiation Monitor, \citep[BERM;][]{Pinto2022}, and near-Earth observations particle observations from the Energetic and Relativistic Nuclei and Electron experiment \citep[ERNE;][]{Torsti1999} and the Electron Proton Helium INstrument \citep[EPHIN;][]{Muller1995} on board SOHO, as well as data from the Solar and Galactic Proton Sensor (SGPS) of the Space Environment In Situ Suite on board GOES \citep[SEISS;][]{Kress2020}. We further present measurements of E$>$150 MeV protons which have propagated through the Mars atmosphere, recorded  on board the Mars Science Laboratory (MSL) by the Radiation Assessment Detector (RAD) \citep{Hassler2012, Guo2021, Papaioannou2019}.

To investigate the evolution of the CME and the shock wave observed in the extreme ultraviolet (EUV) and white light (WL), we use remote-sensing observations of the solar corona provided by the Atmospheric Imaging Assembly \citep[AIA;][]{Lemen2012} on board SDO, the C2 and C3 of the Large Angle and Spectrometric COronagraph \citep[LASCO;][]{Brueckner1995} on board SOHO, and the  Extreme Ultraviolet Imager \citep[EUI;][]{Wuelser2004} COR1 and COR2 coronagraphs, which are part of the Sun Earth Connection Coronal and Heliospheric Investigation \citep[SECCHI;][]{Howard2008} instrument suite on board STEREO.

\begin{figure*}[h!]
\centering
\includegraphics[width=\textwidth]{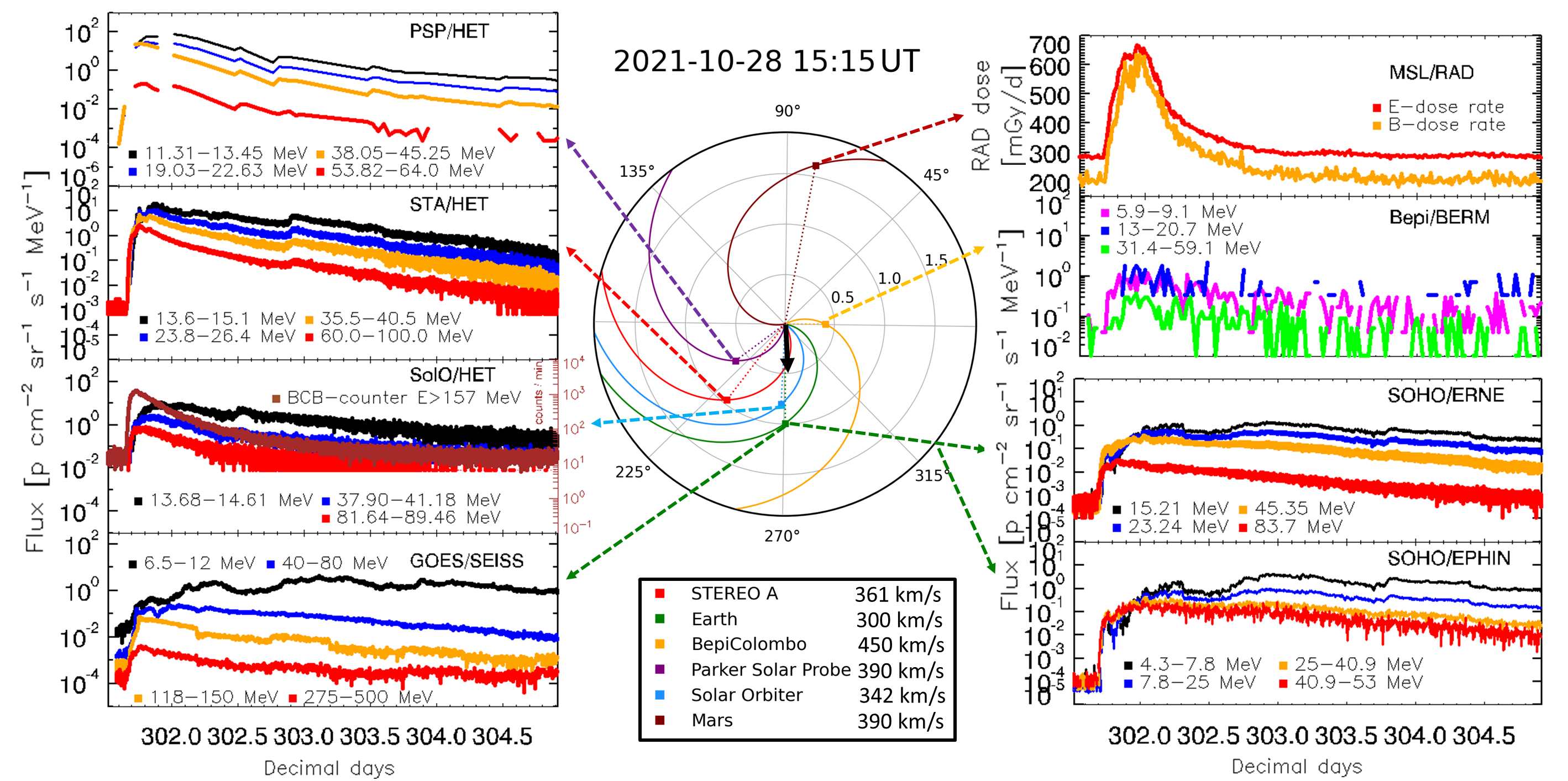}
\caption{ View of the equatorial plane with the location of planets and spacecraft on 28 October 2021, and the SEP recordings during the solar event. The middle panel shows the heliographic equatorial plane from the north, in the Stonyhurst coordinate system, and the location of different spacecraft (coloured squares) at 15:15~UT.  The Parker spirals from the Sun to each spacecraft are shown. The measured (or inferred) solar wind speed used for each observer is provided in the bottom part of the middle panel. The large coloured arrows point to the in situ measurements from each observer shown in the left and right multi-panels. In each panel are shown the measurements of energetic protons. In particular the recordings from PSP/HET (11.31-64.0 Mev), STA/HET (13.6-100 MeV), Solo/HET (13.68-89.46 MeV \& E$>$157 MeV), GOES/SEISS (6.5-500 MeV), SOHO/EPHIN (4.3-53 MeV), SOHO/ERNE (15.21-68.7 MeV), Bepi/BERM (5.9-59.1 MeV) and MSL/RAD dose rates are presented (anti-clockwise from top left). {The black solid arrow denotes the propagation direction of the apex.} } 
\label{fig:sc_pos}
\end{figure*}

\section{Overview of the multi-spacecraft SEP event}

On 28 October 2021 the first GLE event (GLE73) of solar cycle 25 (SC25) was observed by several neutron monitors (NMs) around the Earth \citep[see][]{Papaioannou2022}. High-energy protons were also observed by widely separated spacecraft, associated with GLE73, making this event the first multi-spacecraft high-energy SEP event of SC25. Figure~\ref{fig:sc_pos} shows the positions of various spacecraft in the inner heliosphere and the Parker spirals connecting at each spacecraft. At the time of GLE73, SolO, STEREO-A, and PSP were trailing Earth by -3$^\circ$, -38$^\circ$, and -54$^\circ$, respectively, while BepiColombo was leading Earth by 90$^\circ$ and Mars by 169$^\circ$, as we show in Fig.~\ref{fig:sc_pos}. Moreover, SolO was located at a radial distance of 0.80~au, PSP at 0.62~au, STA at 0.96~au, BepiColombo at 0.41~au, and Mars at 1.61~au.

\cite{Papaioannou2022} showed that GLE73 was associated with an X1.0 class flare, starting at 15:17~UT and peaking at 15:35~UT. The source active region NOAA AR12887 was located at W02S26 (in HGS system at 15:20~UT) as observed from Earth's viewpoint. Radio observations were also very rich for this event and show that solar energetic electrons were accelerated and released in different regions in the solar corona. From metric to kilometric wavelengths (radio domain) type III, type II, and IV radio bursts were observed in association to the event \cite[see details in][]{Klein2022}. The radio observations show three different groups of decametric to kilometric type III bursts that mark different episodes of the significant release of energetic electrons to interplanetary space. According to the analysis of \cite{Klein2022}, the first group of type III radio bursts was probably produced by the observed shock wave, highlighting the possibility that a strong shock wave formed in the low corona from the early phases of the event and accelerated SEPs. {\cite{Klein2022} show that there is also a type IV radio burst that starts in the early phase of the event, which indicates trapped electrons inside the flux-rope.} Additionally, high-energy $\gamma$-rays were observed by the FERMI-Large Area Telescope.\footnote{\url{https://hesperia.gsfc.nasa.gov/fermi/lat/qlook/lat_events.txt}}

\section{Observations and  data analysis} \label{sec:ob}

\subsection{EUV and white-light observations} \label{sec:rs}

\begin{figure*}[h!]
\centering
\includegraphics[width=\textwidth]{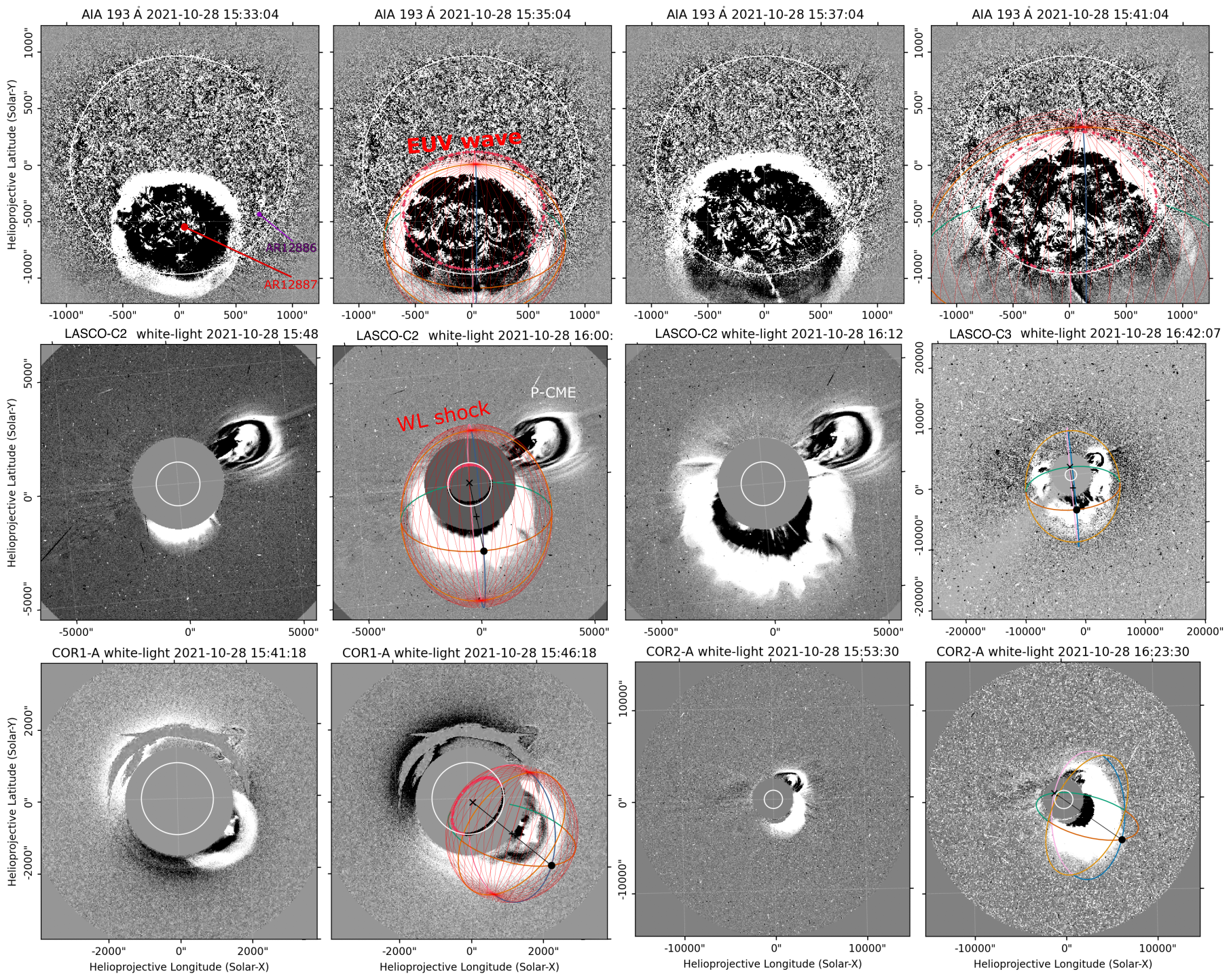}
\caption{Selected snapshots of EUV and WL coronagraphic observations during the 28 {October 2021} solar event. The top row  shows running-difference images from SDO/AIA at 193~\AA, the middle and bottom rows show coronagraphic running-difference images from SOHO/LASCO and STEREO-A, respectively. The EUV wave, the CME, and the WL shock is clearly visible in these panels {and they are labelled in some frames. A narrow streamer blowout CME above the west limb is also shown,  labelled `P-CME'.} The ellipsoid fitted to the shock front during the 3D shock reconstruction is over-plotted to selected frames.} 
\label{fig:remote_sensing}
\end{figure*}

The 28 October 2021 eruptive event was observed with remote-sensing instruments from two vantage points, namely Earth and STEREO-A spacecraft, which  were separated by 38$^\circ$ \citep[see][to reconstruct the CME dynamics based on Earth and STEREO-A remote sensing observations]{Li2022}. These observations are summarized in Fig.~\ref{fig:remote_sensing} where we show a sequence of images in EUV from AIA, and in WL from SOHO/LASCO and STEREO-A/COR1 and COR2. In the low corona an EUV wave was observed by SDO/AIA and STEREO-A/EUVI. In panel~(a) of Fig.~\ref{fig:remote_sensing} we show AIA observations during the evolution of the EUV wave in the low corona. The EUV wave was observed as a bright and coherent propagating front that quickly evolved as a global wave from the Earth's viewpoint. It seems that the  EUV wave propagated outwards from the parent active region (12887) in almost all directions, and continued its expansion nearly uninterrupted for a long time. The EUV wave expanded at an average speed parallel to the solar surface of $\sim$650$\pm$100~km/s, depending on the direction of expansion. A detailed kinematical analysis of the EUV wave that performed by \cite{Hou2022} showed that the wave propagates at an initial speed of 600 to 720 km/s. Their results also suggest that the EUV wave is a fast-mode magnetohydrodynamic (MHD) wave or shock driven by the expansion of the associated CME.

The event was also associated with a CME and WL shock that was observed higher in the corona by the SOHO/LASCO and STEREO-A coronagraphs. Both viewpoints observed a broad CME with a clear bright front surrounding it. {In the middle and bottom rows} of Fig.~\ref{fig:remote_sensing}, we show observations of the CME and the shock from SOHO/LASCO and STEREO/COR1-2-A. According to \citep{Kwon2014}, fast and wide CMEs can perturb the entire corona and can drive very wide pressure waves that in many cases can   encompass the whole corona \citep[e.g.][]{Kwon2017} appearing as the halo CME signatures at most of the observing viewpoints situated around the Sun. These disturbances driven by powerful CMEs can steepen into shock waves if they propagate faster than the local fast-magnetosonic speed in the corona. This is very close to what the  SOHO/LASCO and STEREO-A coronagraphs observed for this powerful event: a wide CME and CME-driven shock wave that propagate fast in the solar corona. {Figure~\ref{fig:remote_sensing} also shows a narrow streamer blowout CME above the west limb that exhibited a slow evolution during and after the primary eruption.} \cite{Papaioannou2022} showed that the plane-of-sky {speed of the GLE73 associated CME} at the leading edge was around 1240~km/s, whereas at the same direction the WL shock had a speed at the plane-of-sky of about 1640~km/s. We present more details on the shock kinematics in Section~\ref{sec:shock_reco} where we reconstruct the shock wave using multi-viewpoint observations and determine the shock speed in 3D.

\subsection{Shock kinematics} \label{sec:shock_reco}

To determine the position and kinematics of the pressure or shock wave more accurately as it propagates in the corona, we reconstructed its 3D structure using observations from different vantage points. To perform the 3D reconstruction we used \textit{PyThea}. This is a software package written in the  Python language that can be used to reconstruct the 3D structure of CMEs and shock waves \citep{Kouloumvakos2022PyThea}. The tool is available online from {GitHub} and Zenodo.\footnote{\url{https://doi.org/10.5281/zenodo.5713659}} For the 3D reconstruction of the shock wave front we used an ellipsoid geometrical model and near-simultaneous multi-viewpoint observations of the shock in EUV and WL. We took advantage of the two viewpoints provided by \textit{STEREO-A} and the near-Earth spacecraft (i.e. \textit{SOHO} and \textit{SDO}). {The ellipsoid geometrical model has been widely used to model the global large-scale structure of propagating shocks in the solar corona and can be an acceptable approximation for some events that do not exhibit non-spherical or significantly corrugated geometry.}

The ellipsoid model is defined from three positional parameters that adjust the longitude, latitude, and height of the  centre and three geometrical parameters that adjust the length of the three semi-axes. During the reconstruction process we adjust the free parameters of the ellipsoid model to achieve the best visual fit of the model to the observations of the two viewpoints. In Fig.~\ref{fig:remote_sensing} we show the wireframe of the geometrical model overlaid in the remote-sensing EUV and WL coronagraph images. This wireframe depicts the front of the reconstructed shock wave in each image and viewpoint. 

\begin{figure*}[h!]
\centering
\includegraphics[width=0.90\textwidth]{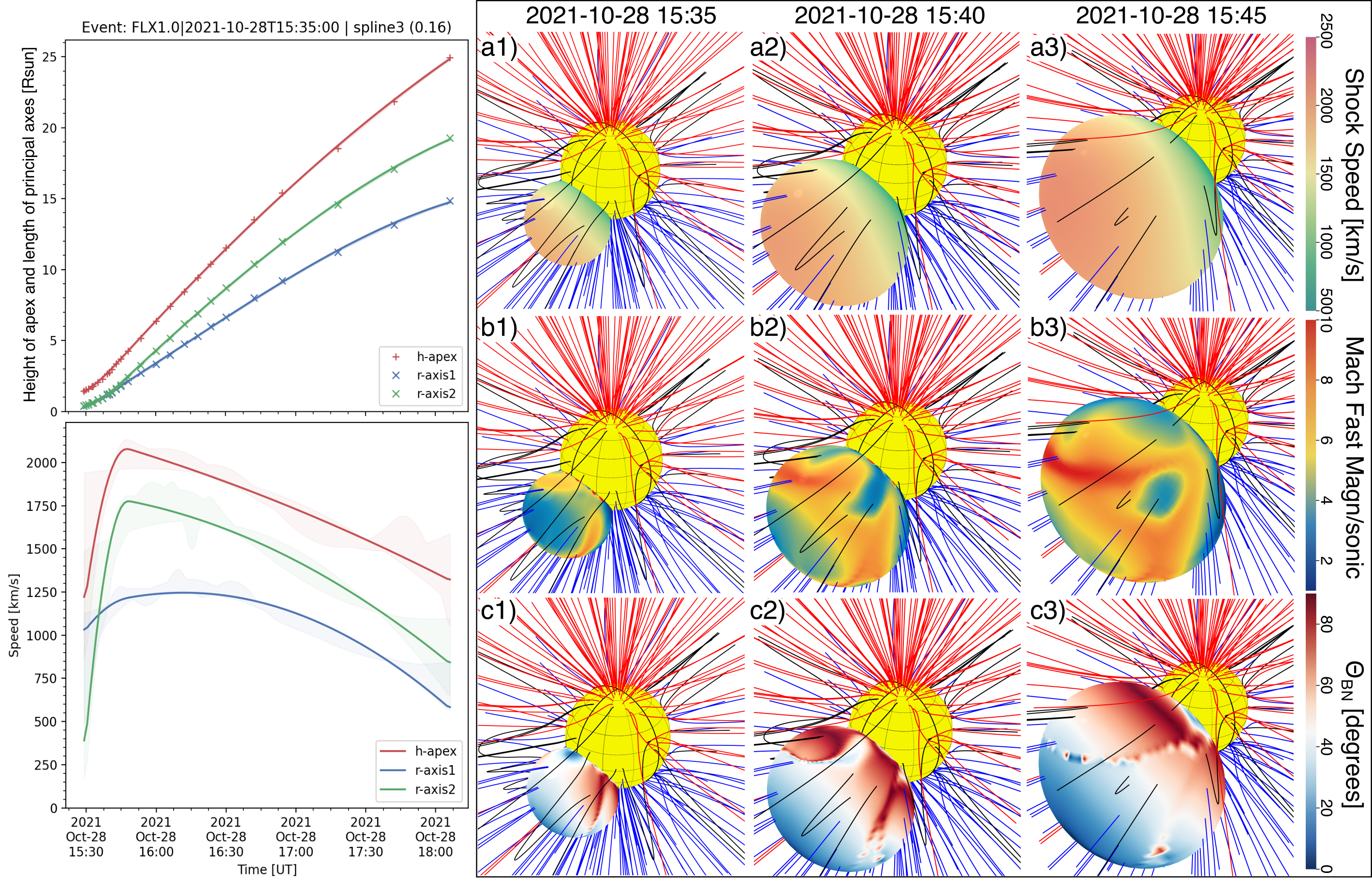}
\caption{Results from the  3D reconstruction and modelling of the shock. The two panels on the left present the {kinematics of the reconstructed shock} using the geometrical ellipsoid model. The top (bottom) panel shows the height (speed) of the shock apex measured from the Sun centre and the lengths (speeds) of the two semi-principal axes of the model. The multi-panel on the right shows selected snapshots of the modelled shock wave parameters  in 3D plotted along the reconstructed pressure wavefront surface. Row (a) shows the shock speed, (b) {the fast-magnetosonic Mach number}, and (c) the shock geometry ($\Theta_{BN}$ angle). The Sun is plotted to scale (yellow sphere) and {the open (red or blue depending on the polarity) and closed (black) coronal field lines traced from the MAS model} are also shown.} 
\label{fig:shock3d}
\end{figure*}

From the 3D reconstruction, we determine the position and kinematics of the shock wave. The 3D reconstruction technique minimizes the projection effects using different viewpoints and allows the shock kinematics to be calculated more accurately. The main direction of propagation for the shock wave apex is found to be at an average longitude  of 2$^\circ$ and latitude of $\sim$24$^\circ$ in the Stonyhurst heliographic coordinate
system. In Fig.~\ref{fig:shock3d}, we show the kinematics of the reconstructed shock wave. Our analysis shows that the shock wave propagated and expanded fast in both the radial and lateral directions. At the shock apex we find a maximum propagation speed of $\sim$2075$\pm$50~km/s and at the shock flanks an expansion speed of $\sim$1500$\pm$75~km/s, which  is an average of the maximum speed of the shock flanks in two different directions. We find a strong acceleration phase for the first 15 minutes after the shock initiation and until 15:48~UT when the speed at the shock apex reached the maximum value. We also find that there is a strong lateral overexpansion of the shock flank in the north--south direction with a maximum speed of $\sim$1800~km/s compared to the east--west direction where the shock flank expands with a maximum speed of $\sim$1200~km/s. This phase starts   at around 15:40~UT and lasts probably well after the end of our shock modelling. Comparing the maximum expansion speed at the two locations at the shock flanks we find that in the north--south direction the shock expands about 1.35 times faster than in the east-west direction. {After this phase the shock starts to decelerate as it progressively evolves as a freely propagating blast shock wave since it probably detaches from its driver, the CME,  which propagates significantly more slowly.}

\subsection{Shock modelling and parameters in 3D} \label{sec:shock_model}

Using the results of the shock kinematics from the 3D reconstruction and  MHD parameters of the background solar corona, we estimated the shock parameters in 3D \citep[see][for further details of the shock modelling]{Rouillard2016, Kouloumvakos2019}. 
{For the MHD parameters of the background corona, we utilized data from the Magnetohydrodynamic Algorithm outside a Sphere \citep[MAS;][]{Lionello2009} thermodynamic model which includes realistic energy equations, radiative losses, and parametrized coronal heating, and it accounts for thermal conduction parallel to the magnetic field. By incorporating these detailed thermodynamic effects, the MAS thermodynamic model provides more precise estimates of plasma density and temperature in the corona \citep{Riley2011}, from the solar surface to 30~R$_\sun$, which is the outer boundary of the model. As the inner boundary condition of the magnetic field the model utilizes photospheric magnetograms from SDO. In this study, we used the high-resolution MAS data cubes from Carrington rotation 2250 provided by  Predictive Science  Inc.\footnote{\url{https://www.predsci.com/}}}

In Fig.~\ref{fig:shock3d} a--c, we show selected snapshots of the modelled shock wave parameters in 3D, plotted along the reconstructed pressure wavefront surface. {Row (a) presents the shock wave's 3D expansion speed. The shock is faster at the apex and slower at the flanks. Row (b)  shows the fast-magnetosonic Mach number at the wavefront surface. We see that there are multiple regions where the Mach number is very high ($>>$4) suggesting that strong shock regions probably formed in the low corona during the event. These regions in row (b) are mainly located close to the neutral line of the heliospheric current sheet where the fast-magnetosonic speed is low \citep[see][]{Rouillard2016}. In Row (c), we present the shock geometry. Near the apex the $\Theta_{BN}$ angle is mostly oblique to quasi-parallel ($\Theta_{BN}$<45$^\circ$), whereas at the flanks the $\Theta_{BN}$ angle is {mostly oblique to quasi-perpendicular} ($\Theta_{BN}$>45$^\circ$).}

\begin{figure*}[h!]
\centering
    \includegraphics[width=0.85\textwidth]{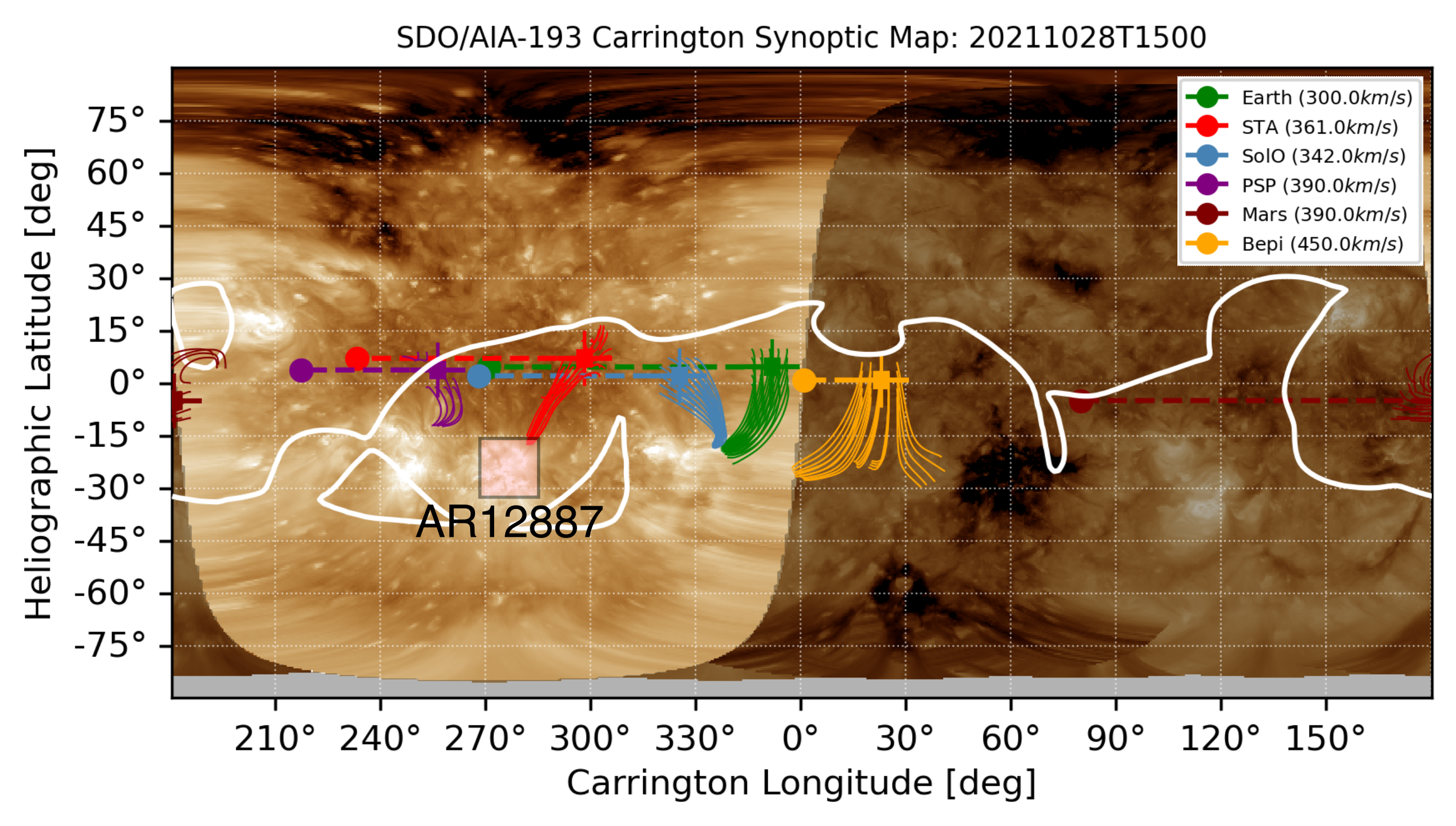}
\caption{Magnetic connectivity of the observers to the solar surface on 28 {October 2021 at 15:00~UT, from the field line tracing using the magnetic field data from the MAS model.} The observers' position is projected to the map and depicted with coloured circles. The Parker spirals connecting the observers to the low corona are shown with the dashed coloured lines and the footpoint of the spiral at 2.5~$R_\sun$ is shown with the coloured squares. The traced magnetic field lines are depicted with the coloured lines. The neutral line at $2.5~R_\sun$ is shown with the white line. The background map is constructed using SDO/AIA images at 193~\AA.} 
\label{fig:MAS_connect}
\end{figure*}

Using the result of the shock model we calculated the evolution of the shock's fast-magnetosonic Mach number ($M_{fm}$) at the field lines connected to each observer. {First, for each spacecraft that observed the SEP event (see Fig.~\ref{fig:sc_pos}), we derived the magnetic connectivity of the observers to the solar surface. For the low-coronal part ($<$30~R$_\sun$) we used the magnetic field data from the MAS MHD model, whereas for the interplanetary magnetic field we assumed a Parker spiral derived using solar wind speed measurements near the time of the SEP event. In the case of BepiColombo, for which there are no solar wind speed measurements, we assumed a value of 450 km/s. For each observer we utilized the Parker spirals to determine the location of the footpoints at 2.5~R$_\sun$.  Subsequently, we performed a field line tracing at MAS data starting from the location of the footpoint, to determine the magnetic  connectivity of each observer in the corona. For this connectivity analysis we assumed an uncertainty of five degrees, which is at the same scale with the size of solar supergranules. The field line tracing is performed within a region spanning an angular extent of five degrees. In general, the state of interplanetary medium can induce a greater uncertainty in the connectivity estimates if the preceding events are present in the interplanetary medium. Nonetheless, upon an examination of COR2 and LASCO-C2 images, no noteworthy preceding CMEs were identified that could substantially disrupt the interplanetary space.}

{Figure~\ref{fig:MAS_connect} shows the results from the magnetic connectivity analysis projected to a Carrington (CR) map constructed using EUV images at 193~\AA\ from SDO/AIA. The magnetic connectivity of the different spacecraft is depicted with the coloured field lines. For most observers, the magnetic field lines seem to diverge many degrees away from the connection points of the Parker spirals. The coronal connectivity in this case can be very broad since the field lines connected to the observers extend more than $10^\circ$ in heliospheric longitude and latitude. This seems to be the case for most of the observers. More specifically, from the magnetic connectivity analysis we find that except for BepiColombo all the other observers were magnetically connected to the visible disk (see Fig.~\ref{fig:MAS_connect}). STEREO-A has the best magnetic connectivity to AR12887 and PSP is also closely connected, whereas for SolO and Earth the connection is more distant from AR12886 ($>$60$^\circ$) and closer to AR12886, with Carrington longitude (CRLN) ranging from $320^\circ$ to $350^\circ$. Mars was also connected far from the parent AR; its field lines are located above the east limb, at CRLN$\sim180^\circ$. BepiColombo has the most distant connectivity from AR12887, and it was partially magnetically connected to the periphery of a small coronal hole that was located at the far side of the Sun. In Section~\ref{subsec:assesment_connect} we assess the connectivity estimates using different models. This analysis showed qualitatively similar results for most of the observers.}

\begin{figure}[h!]
\centering
\includegraphics[width=0.48\textwidth]{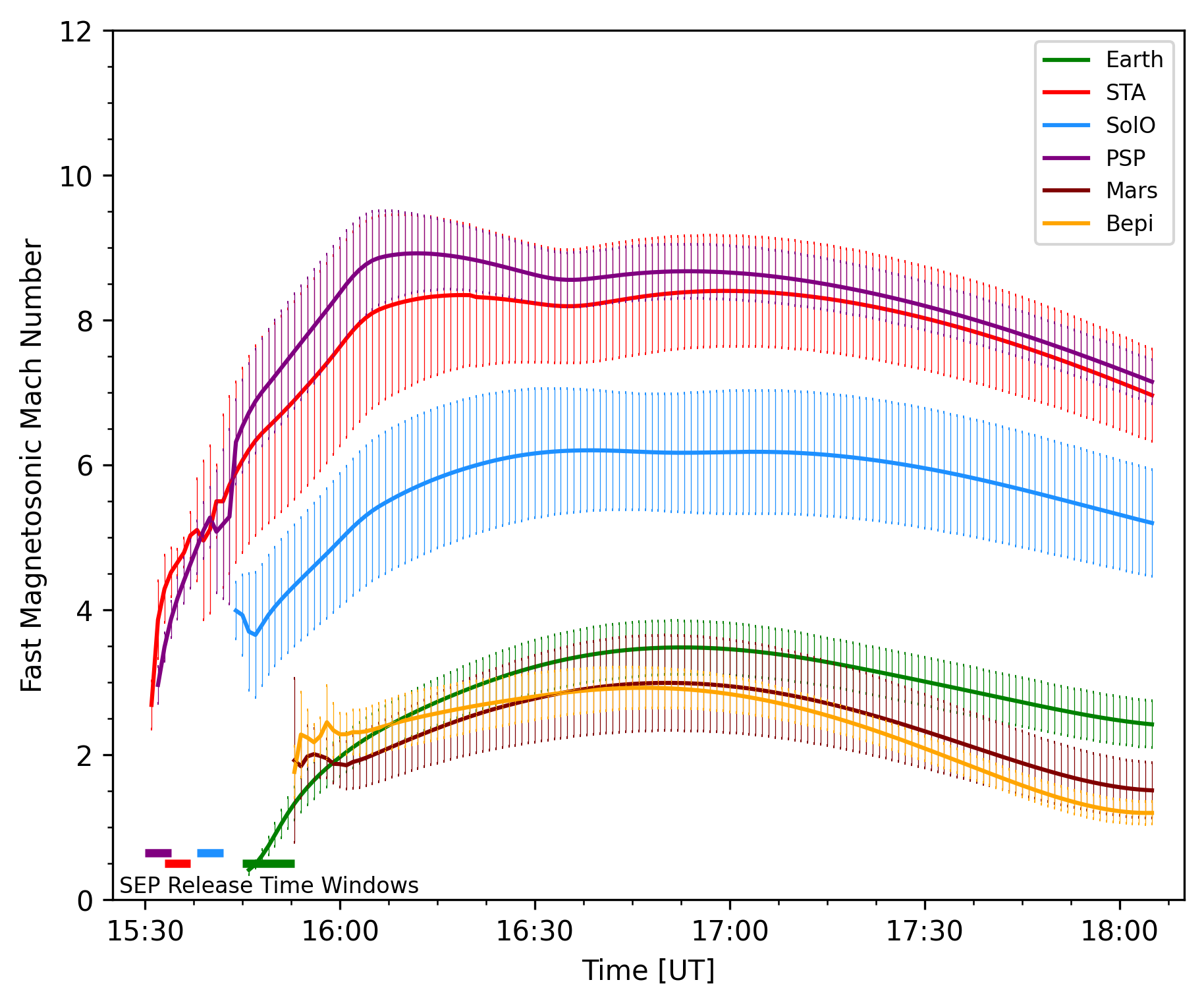}
\caption{Temporal evolution of the mean fast-magnetosonic Mach number at the field lines connected to each spacecraft, {using the connectivity from MAS. The solid coloured lines depict the mean value of $M_{fm}$ and the vertical error bars are the one-sigma values. The horizontal bars depict the SEP release time windows that are determined from the VDA, for each observer.} } 
\label{fig:shParam}
\end{figure}

{In Fig.~\ref{fig:shParam} we show the evolution of the shock $M_{fm}$ at the field lines connected to each observer. We used the results from the connectivity analysis from the MAS data and the shock 3D model to infer the shock parameters at the field lines connected to each observer. With the solid lines we depict the average values of $M_{fm}$ along the field lines connected to the observers. The error bars are the one-sigma values calculated from the standard deviation of the parameter.} From the results of this analysis, we find that PSP was connected to a strong shock region with the peak {$\mathrm{M_{fm}\sim8.9\pm0.5}$.} The $M_{fm}$ for the shock region connected to PSP remains very high until the end of the shock modelling. We find a {mean value of $\overline{M}_{fm}$=7.9$\pm$0.4} during the time interval that we model. The good proximity of the connection points to the heliospheric current sheet may have played a significant role in the formation of a strong shock with high Mach numbers. Furthermore, STEREO-A was connected {to strong shock regions with a peak $M_{fm}$=8.4$\pm$0.8 and mean value $\overline{M}_{fm}$=7.6$\pm$0.9,} and SolO was connected to moderate strength shock regions with a peak {$M_{fm}$=6.2$\pm$0.8 ($\overline{M}_{fm}$=5.7$\pm$0.8).} On the other hand, all the other observers, Earth, Mars, and Bepi, were connected to relatively weak shock regions. More specifically, {for Earth we find a peak $M_{fm}$=3.4$\pm$0.4 ($\overline{M}_{fm}$=2.8$\pm$0.3), for Bepi $M_{fm}$=2.9$\pm$0.3 ($\overline{M}_{fm}$=2.3$\pm$0.3), and for Mars $M_{fm}$=2.6$\pm$0.4 ($\overline{M}_{fm}$=2.4$\pm$0.6)}.

\subsection{SEP observations} \label{subsec:overview}



This high-energy SEP event was clearly observed {by multiple spacecraft as shown in Fig.~\ref{fig:sc_pos},} making it a multi-spacecraft event. What is more, Mars,  which was located almost diametrically opposite from Earth during the event (see Fig.~\ref{fig:sc_pos}), recorded the event, and thus the spread of GLE73 was almost 360$^\circ$ around the Sun. The time evolution of the event was recorded {by many instruments, as shown in Fig.~\ref{fig:sc_pos}. Specifically, we show the  PSP/EPI-Hi/HETA energetic proton measurements spanning from 11.31 to 64~MeV, STEREO-A (STA)/HET at energies covering from 13.1 to 100~MeV,  SolO/HET measurements from 13.68 to 89.46~MeV and  BCB-counter (with E>157~MeV; in counts/min; \cite{vonForstern2021}), and GOES/SEISS differential data in the range  6.5\,--\,500~MeV.} Moreover, SOHO/ERNE recorded the event over a large energy range (from 10 to 100 MeV), and the channels with effective energies from 15.4 to 57.4 MeV are also depicted in Fig.~\ref{fig:sc_pos}. SOHO/EPHIN also recorded the event at low energies (from 4.3 to 53~MeV), as presented in Fig.~\ref{fig:sc_pos}. The Bepi/BERM 30 min averaged data show an increase at the  5.9\,--\,59.1~MeV range. Finally, RAD on board MSL on the surface of Mars recorded a distinguishable increase at both E- and B-dose rates. The required energy for the initiation of a proton triggering a GLE recorded by RAD located in Gale crater on Mars was $\sim$ E$>$150~MeV \citep{Guo2018}. {Moreover, detailed reconstructions can further enhance the detection capabilities of SOHO/EPHIN and SolO/HET to energies spanning from $\sim$49 to $\sim$600~MeV \citep[see details in][]{Kuhl2015, Kuhl2019} for the former and to protons of $\sim$300 MeV and particles up to $\sim$900 MeV for the latter.} Both SOHO/EPHIN and Solo/HET have clearly measured GLE73 (see further details in Appendix \ref{appendixA}).

\subsubsection{SEP release times}

High-energy {protons have} a prompt increase in most of the observers (all indicated with a red line in each sub-panel of Fig.~\ref{fig:sc_pos}). STEREO-A/HET (60\,--\,100~MeV) observes the first arriving high-energy protons at 15:54~UT. Unfortunately, there is a data gap in the highest energy channels for PSP/HET (E=58.68 MeV) and the time profile is obtained halfway through the rise time, precluding a determination of the exact onset time. For SolO/HET the very high-energy proton channel (E=300.88 MeV) has an onset time at 15:49~UT. For the near-Earth spacecraft we find that the reconstructed measurements at $\sim$610 MeV by SOHO/EPHIN observed an onset at 15:50~UT, while GOES/P10 (275\,--\,500~MeV) had an onset time at 15:55~UT, and SOHO/ERNE (57.4 MeV) recorded the onset of the event at 16:18~UT. At Earth, \cite{Papaioannou2022} showed an onset of 15:45~UT for the GLE73 event from the recordings of the  south pole neutron monitor (SOPO NM) station, which registered the earliest onset (at $>$430 MeV). Finally Mars/RAD (E$>$150~MeV) measured the start of the event with a relative delay at 16:34~UT and Bepi/BERM recorded a clear increase above background in the 5.9\,--\,9.1~MeV channel\footnote{According to \citet{Pinto2022}, this particular channel for Bepi/BERM (i.e. P$\_$BIN$\_$3; 5.9\,--\,9.1~MeV) is optimal for the identification of protons arriving at the spacecraft since there is no electron contamination present.} at around 17:00~UT (when using 30 min averaged data).    

\begin{figure}[h!]
\centering
\includegraphics[width=0.46\textwidth]{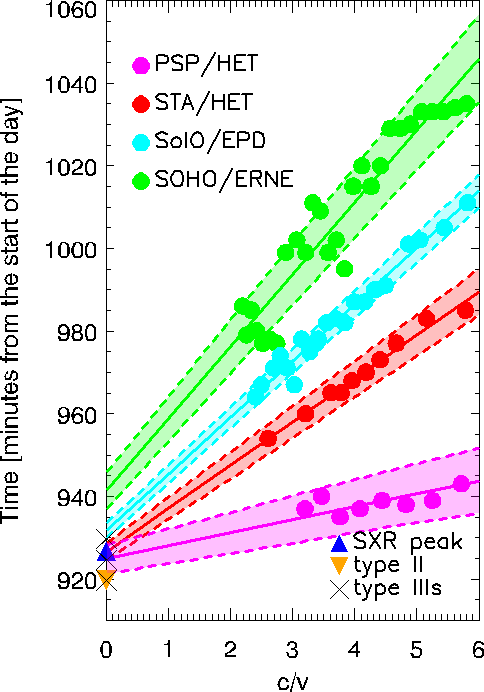}
\caption{VDA results for PSP/HET (magenta circles), STA/HET (red circles), SolO/EPD (cyan circles), and SOHO/ERNE (green circles). The lines depict the obtained linear fits, while the dotted lines and the corresponding shaded area with the same colouring represent the 1$\sigma$ error per fit. The peak of the SXR flare is presented as a blue triangle, the type II burst at m-$\lambda$ with a reverse orange triangle, while the three episodes of type III bursts are represented with a black X symbol. All solar eruptive signatures have been shifted; see text for details.} 
\label{fig:VDA}
\end{figure}


\begin{table*}[!ht]
    \centering
    \caption{SEP onset and estimated release times for the high-energy particles recorded by the various spacecraft.}
    \begin{tabular}{lcccrc}
        \hline
       Observer & Energy & Onset& Path & Release & $M_{fm}$ \\ 
         & (MeV) & time & Length$^{1}$ & time$^{2}$ & \\
         &  & (UT) & (au) & (UT) & \\ \hline
        PSP  & 10.37\,-\,49.35  &  & 0.38$\pm$0.09 & 15:33$\pm$4 & 4.2$\pm$0.8\\ 
        STA  & 13.1\,-\,100 &  & 1.26$\pm$0.06 & 15:35$\pm$2 & 4.8$\pm$0.6 \\
        SolO & 10.98\,-\,104.9 &  & 1.64$\pm$0.04 & 15:40$\pm$2 & 3.9$\pm$0.9 \\ 
        SOHO & 13.36\,-\,107.26  &  & 2.09$\pm$0.12 & 15:49$\pm$4 & 1.2$\pm$0.2 \\
        \hline
        \hline
        SOPO NM$^*$ & $>$430 & 15:45 & 1.28 & $\sim$15:40 & \\
        SolO/HET & 300.88 & 15:49 & 0.91& 15:45 & \\
        SOHO/EPHIN & 610 & 15:50 & 1.28 &  15:45 & \\
        GOES/SEISS & 275\,-\,500 & 15:55 & 1.28 &  15:48 & \\ 
        Mars/RAD & $>$150 & 16:34 & 2.19& 16:06 & \\
        Bepi/BERM & 5.9\,-\,9.1 & $\sim$17:00 & 0.42 & & \\\hline
        \end{tabular}\label{tab:VDA}
        \tablefoot{ $^{1}$: The inferred path lengths ($L$, AU) from the VDA is presented; the calculated path length based on the solar wind speed for each observer is also given for the cases where the TSA was applied. $^{2}$: The anticipated Solar Release Time (SRT) deduced by VDA applied to PSP, STA, SolO, and SOHO (first four rows) and TSA applied to the very high-energy channels of SolO, SOPO NM, SOHO/EPHIN, GOES/SEISS, and Mars/RAD. $^*$: SOPO NM results are from \cite{Papaioannou2022}}
\end{table*} 
 
To evaluate the SEP solar release time (SRT) for each observer and the length of the IMF spiral, $L$, along which the particles travelled, we performed a velocity dispersion analysis (VDA) \citep[see details in][]{Vainio2013}. We applied VDA using the measured onset times which were identified using the Poisson-CUSUM (PCM) method \citep[see e.g.][]{Huttunen2005, Kouloumvakos2015, Paassilta2018} or the $n-\sigma$ criterion (SC) \citep[see e.g.][]{Papaioannou2014a, Papaioannou2014b}, depending on the data used. Not all the methods could be applied successfully for all the spacecraft. We give further details for this in Appendix~\ref{appendixB}. Additionally, time-shifting analysis (TSA) \citep[][]{Vainio2013, Papaioannou2022} {was utilized} to determine the release times in selected data products and energy channels, such as the reconstructed fluxes from SolO/HET. All results are summarized in Table \ref{tab:VDA};    the {upper part} provides the results for the VDA and the lower part from TSA.


Figure~\ref{fig:VDA} shows the SEP onset times as a function of the inverse velocity and the obtained linear fit to the onset time of all four spacecraft that we used to  perform the VDA. For each energy channel we used the mean energy to calculate the inverse velocity. The continuous lines in Fig.~\ref{fig:VDA} depict the linear regression for each case (i.e. PSP/EPI-Hi/HETA, magenta; STA/HET, red;  SolO/EPD, cyan; SOHO/ERNE, green) and the dotted lines with the shaded filled area (in the same colours) demonstrate the   1$\sigma$ (68\% confidence) error of each fit. Additionally, on the vertical axis of the plot in  Fig.~\ref{fig:VDA} we indicate the soft X-ray (SXR) peak time with a blue triangle, the time of the type II burst with a reverse orange triangle, and the three type III episodes as described in \cite{Klein2022} with a black $\rm X$. The results of the VDA are presented in Table \ref{tab:VDA}. From this analysis we find that the earliest release of SEPs was for PSP at 15:33($\pm4~min$)~UT. Then for STA we find a release time at 15:35($\pm2~min$)~UT and for SolO at 15:40($\pm2~min$)~UT. For the near-Earth spacecraft (i.e. SOHO) we find a release time at 15:49($\pm4~min$)~UT. The uncertainties of the obtained SRT for the observers should further take into account the ambiguity of the earlier onset time determination as noted  above. Thus, for STA and PSP the uncertainty of the release times is not less than $\pm$5~min. Additionally, TSA suggests that very high-energy protons recorded in the near-Earth space were released no later than $\sim$11 min after the peak of the flare. 

Using the results from the VDA and the 3D shock model (Sect.~\ref{sec:shock_model}), we determine the shock strength (quantified by the fast-magnetosonic Mach number) at the SEP release time for each observer. Because of the uncertainty in the estimates of the SEP release times, we calculate, for each observer, the mean value of the shock fast-magnetosonic Mach number during the SEP release time window. The connection of the modelled shock is found to be inside the SEP release time window for every observer except for SolO, which {is three minutes later. This discrepancy is probably caused by a combination of the uncertainties in the shock model, the connectivity, and the estimated SEP release times. We calculated $M_{fm}$, but only for the  SolO case,  from the time of the first connection of the modelled shock to the well-connected field lines.} At PSP We find that  strong shock regions were promptly connected to the spacecraft, so {$M_{fm}$=3.2$\pm$0.6 at the SEP release time window; for STA we find 4.7$\pm$0.3, for SolO 3.9, and lastly for SOHO/ERNE  1.3$\pm$0.7}.

\subsubsection{SEP spectral properties}

\begin{figure}[h!]
\centering
\includegraphics[width=0.90\columnwidth]{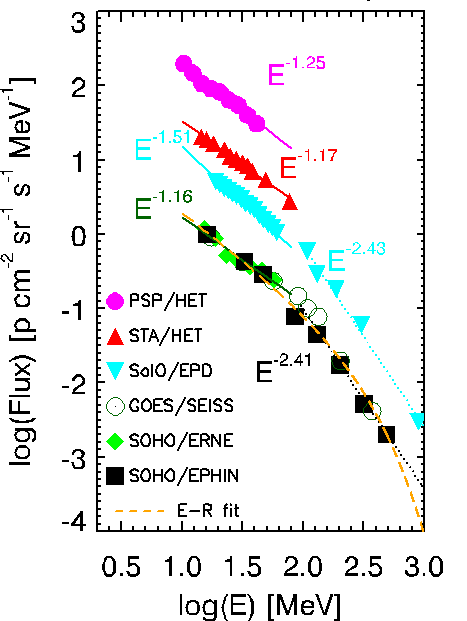}
\caption{Peak proton flux differential spectrum at the interplanetary space measured by PSP/EPI-Hi/HETA (magenta points), STA/HET (red triangles), SolO/EPD (blue triangles), GOES/SEISS (open circles), SOHO/ERNE (green diamonds), and SOHO/EPHIN (black squares). The dotted black and cyan lines indicates the inverse power-law spectra for the higher energy part for Earth (i.e. GOES and SOHO) and SolO, respectively. A slope of (a) $\mathrm{\gamma \sim2.41\pm0.15}$ was obtained from GOES/SEISS and SOHO/EPHIN measurements from 86\,--\,610 MeV (black dotted line) and (b) $\mathrm{\gamma \sim2.43\pm0.08}$  from SolO/EPD from 108\,--\,896 MeV (cyan dotted line). The dashed orange line depicts the obtained E-R fit from Eq~(\ref{eq1}).} 
\label{fig:GOEs_spectra}
\end{figure}

For this high-energy multi-spacecraft SEP event, direct observations of the peak proton flux were obtained by spacecraft measurements in the interplanetary space in a very extended energy range from 15.4 to $\sim$900 MeV. Figure~\ref{fig:GOEs_spectra} provides the differential peak proton spectrum from PSP/HET (magenta points), STA/HET (red triangles), SolO/EPD (blue triangles), GOES/SEISS (open green circles), SOHO/ERNE (green diamonds), and SOHO/EPHIN (black squares). The peak intensities (in units of protons $ \rm cm^{-2} sr^{-1} s^{-1} MeV^{-1}$, 5 min averages) in their prompt component were identified as the maximum intensity observed shortly after the onset of the event {in situ,} excluding the energetic storm particles. For some events the maximum intensity in the prompt component is observed as a plateau in the time-intensity profile. In these cases the peak intensity is taken as the maximum value of the intensity plateau \citep[see details in][]{Papaioannou2023}.
As can be seen, the energy spectrum slope for PSP/HET (spanning from 10.37 to  41.50 MeV) and SolO/EPD (spanning from 17.52 to 85.0 MeV) present an inverse power-law dependence with an exponent of $\gamma$ = 1.25$\pm$0.04 and $\gamma$ = 1.51$\pm$0.05, respectively. Moreover, for STA/HET (spanning from 13.6 to 100 MeV) the obtained $\gamma$ is 1.17$\pm$0.04, in the range 13.6--100~MeV.

Since the low-energy measurements of GOES/SEISS seemed to be contaminated (possibly due to the penetration of high-energy particles or electrons in the lower energy channels; see \cite{Papaioannou2022}) data above 90 MeV were used in the spectra analysis. Therefore, in the low-energy part at near-Earth space, SOHO/ERNE and SOHO/EPHIN recordings, covering 15.4\,--\,57.4 MeV, were used and an inverse power law with an exponent of $\gamma_{low}$ = 1.16$\pm$0.07 was obtained. {All lower energy spectra are depicted as solid lines, following the colour-coding of each observer employed.} In addition, using GOES/SEISS and SOHO/EPHIN measurements, in the range 86-610 MeV an inverse power law with an exponent $\gamma_{up}$ = 2.41$\pm$0.15 {was found. Additionally, the very high-energy recordings from SolO/HET were fitted with an inverse power law leading to a comparable exponent of $\gamma$ = 2.43$\pm$0.08. Both fits for high-energy particles are presented in Fig.\ref{fig:GOEs_spectra}; for Earth (i.e. GOES and SOHO) and SolO}. The low- to high-energy part appears to be separated at $\sim$ 90 MeV. This is in agreement with \cite{Zhang2022} who utilized only GOES/SEISS measurements and showed that a double power law does exist for the peak proton flux spectrum (their Figure 5), with the low-energy part ($<$80 MeV; according to these authors) leading to an exponent of 0.90$\pm$0.02 and the high-energy part ($\geq$80 MeV; based on that paper) having an exponent of 2.51$\pm$0.04. Very high-energy data\footnote{The details of these measurements are presented in Appendix \ref{appendixA}.} from SOHO/EPHIN and SolO/HET are additionally shown in Fig.~\ref{fig:GOEs_spectra}.  

The presented results corroborate with an expectation of two power laws, demonstrating that the peak proton flux energy spectrum in the low-energy range is harder than that in the high-energy range. The double power law of the peak proton flux energy spectrum has been explained by shock acceleration \citep[see e.g.][]{Tylka2001, Tylka2005} or transport effects \citep[][]{LiLee2015}. Moreover, \cite{Kiselev2022} suggested that particles of different energy ranges may have different acceleration mechanisms (one assumed to be CME or shock-related and the other flare-related) with distinct signatures in the low- and high-energy parts, implying that the shock-related contribution dominates the low-energy particles. The presence of two different accelerators complicates the identification of the sources of SEPs. {Diffusive shock acceleration (DSA) predicts that for low-energy particles a planar shock will lead to a power-law energy spectrum \citep[see e.g.][]{Ellison1985}. For higher energies an exponential rollover will be present, emerging for example  due to  particle losses, limited acceleration time, and adiabatic cooling \citep[see e.g.][and references therein]{GLi2022, Yu2022}. Thus, according to the treatment presented in \cite{Ellison1985}, a differential spectrum in the form of an inverse power law with an exponential  rollover (hereafter E-R), Eq.~(\ref{eq1}) can be obtained,} 


\begin{equation}
    \frac{dJ}{dE} = KE^{-\gamma} \exp\left(-\frac{E}{E_0}\right)
\label{eq1}
,\end{equation}
where $J$ is the intensity; $E$ is the kinetic energy/nucleon; and  $K$, $E_0$, and $\gamma$ are constants \citep{Mewaldt2012, Kiselev2022}. {As noted  above, this spectrum has a power-law shape at low energies, as expected from shock acceleration, with an exponential rollover at high energies, presumably determined by the finite radius of the shock or the time available for accelerating particles to high energy \citep{Mewaldt2005}}. 

Combining the measurements from SOHO/ERNE, SOHO/EPHIN, and GOES/SEISS (excluding low-energy particles from GOES) an E-R fit (Eq.~(\ref{eq1})) was applied (orange dashed line in Fig.~\ref{fig:GOEs_spectra}). The obtained rollover energy is $E_0$ = 187.10 MeV, $K$ = 28.13, and $\gamma$ = 1.16$\pm$0.05. The E-R spectra seems to be in good agreement with the measurements from 15.4 up to 610 MeV. 

\subsection{SEP composition properties}

\begin{figure}[h!]
\centering
\includegraphics[width=0.94\columnwidth]{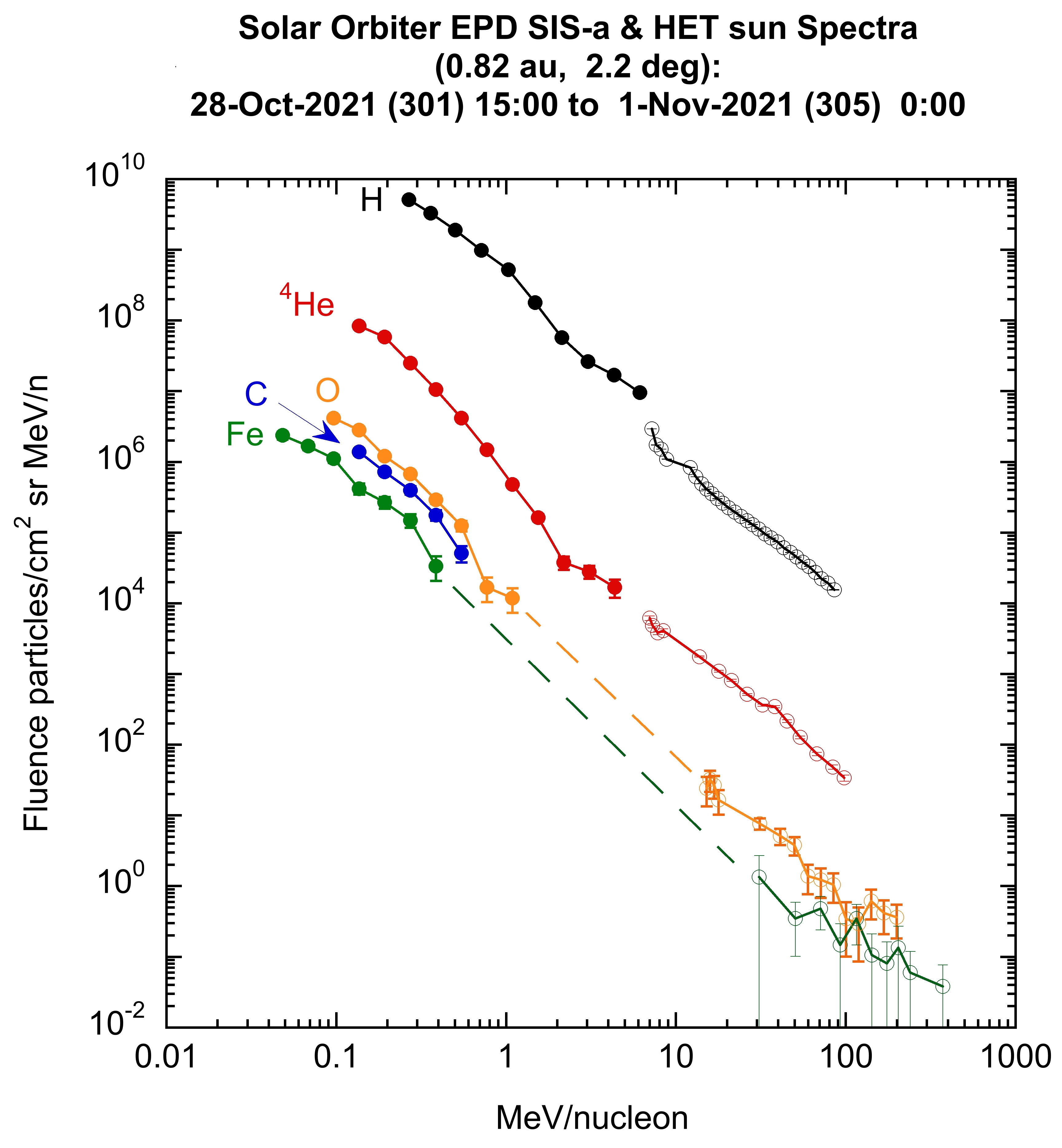}
\includegraphics[width=0.94\columnwidth]{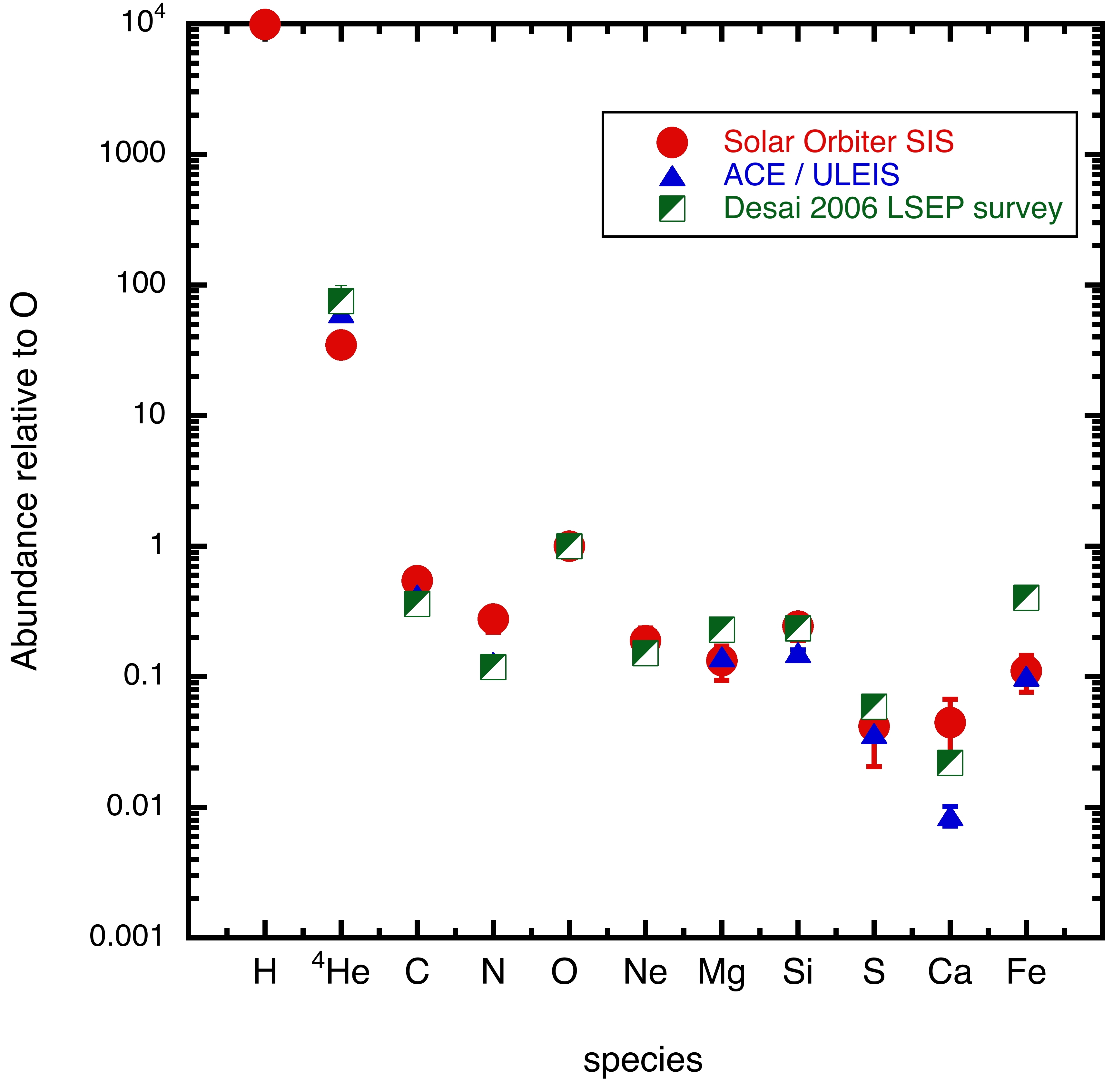}
\caption{ Spectra and relative abundance over the entire event. Top: Differential fluence spectra for major species. The dashed lines connect gaps in instrumental energy coverage and are only  to guide the eye. Bottom: Relative abundance of ions measured over the energy range 320\,--\,450 keV/nucleon from SOLO/SIS, ACE/ULEIS, and the survey of \cite{Desai2006}. }
\label{fig:SOLO_SIS}
\end{figure}

Figure~\ref{fig:SOLO_SIS} (top panel) shows the energy spectra from the SolO/EPD SIS and HET sunward-looking telescopes summed over the entire event (from 28 October 2021 15:00~UT to 1 November 2021 00:00~UT). The spectra are roughly power laws over the range above $\sim$0.3 MeV/nucleon, with the heavier ions showing rollover below $\sim$0.1 MeV/nucleon. Above a few hundred keV/nucleon there are no obvious steepening or `breaks' in the spectra, which are often observed \citep{Cohen2005, Desai2016b}. This is probably due to the increase in intensities below 2~MeV/nucleon when the shock passed by SolO late on 30 Oct, and thus causing the low-energy portion of the spectra to  steepen. The event averaged proton spectral slope is -1.71 over the range 0.3\,--\,1.0~MeV and only slightly steeper (-1.83) over the range 20\,--\,90~MeV. These are steeper than the 10\,--\,100 MeV slopes shown for the peak proton flux spectra in Fig.~\ref{fig:GOEs_spectra}.

The average abundances of 320\,--\,450keV/nucleon ion species normalized to O are shown in the bottom panel of Fig.~\ref{fig:SOLO_SIS}, averaged over the event for both Solar Orbiter/SIS and ACE/Ultra Low Energy Isotope Spectrometer \citep[ULEIS;][]{Mason1998}. The panel also shows the average from the broad survey of large SEP events by \cite{Desai2006}, The survey of Desai et al. excluded SEP events with energetic storm particle increases. The result from this panel showing that the GLE73 had a composition very similar to large solar particle events, except that the  Fe/O was close to the low end of the distribution of values found in the survey. Additionally, there was no evidence for enhancement of 3He in this event ($^3$He/$^4$He<1\% between 0.5\,--\,2.0~MeV/nucleon). In SOHO/ERNE the  He-to-p ratio at 50 MeV/n is below 1\% and the  Fe-to-O ratio is around 0.2 at similar energies, consistent with observations of SolO/HET.

\subsection{SEP modelling}

The results of the shock modelling analysis (see Sect~\ref{sec:shock_model}) showed that the Earth was magnetically connected to weak shock regions. Although these regions may be able to accelerate protons, the inferred Mach numbers suggest the acceleration efficiency is probably low for these distant magnetically connected observers. To explore whether particle transport may play an important role  towards high-energy SEPs reaching these observers, we carried out 3D test particle simulations of the event. The simulations were performed for proton energies in the range {50\,--\,1000 MeV} and the results compared to high-energy observations by GOES and other spacecraft. The test particle code \citep{Dalla2005} was   previously used to model energetic particle transport in the heliosphere \citep{Marsh2013, Battarbee2018}. It uses a Parker spiral magnetic field and includes the drifts associated with its gradient and curvature. Recently, it has been used to model drift effects along the heliospheric current sheet for other GLE events \citep{Waterfall2022}.
Turbulence is included as pitch angle scattering (which induces some cross-field motion) with a parallel mean free path of 0.3 au assumed; however, no perpendicular diffusion or field line meandering are included in the simulation. While models of these processes do exist, the interplay between turbulence and particle transport close to the HCS is currently unknown and it is difficult at the present time to include both the HCS and perpendicular diffusion and/or field line meandering in our simulations.

\begin{figure}[h!]
\centering
\includegraphics[width=0.50\textwidth]{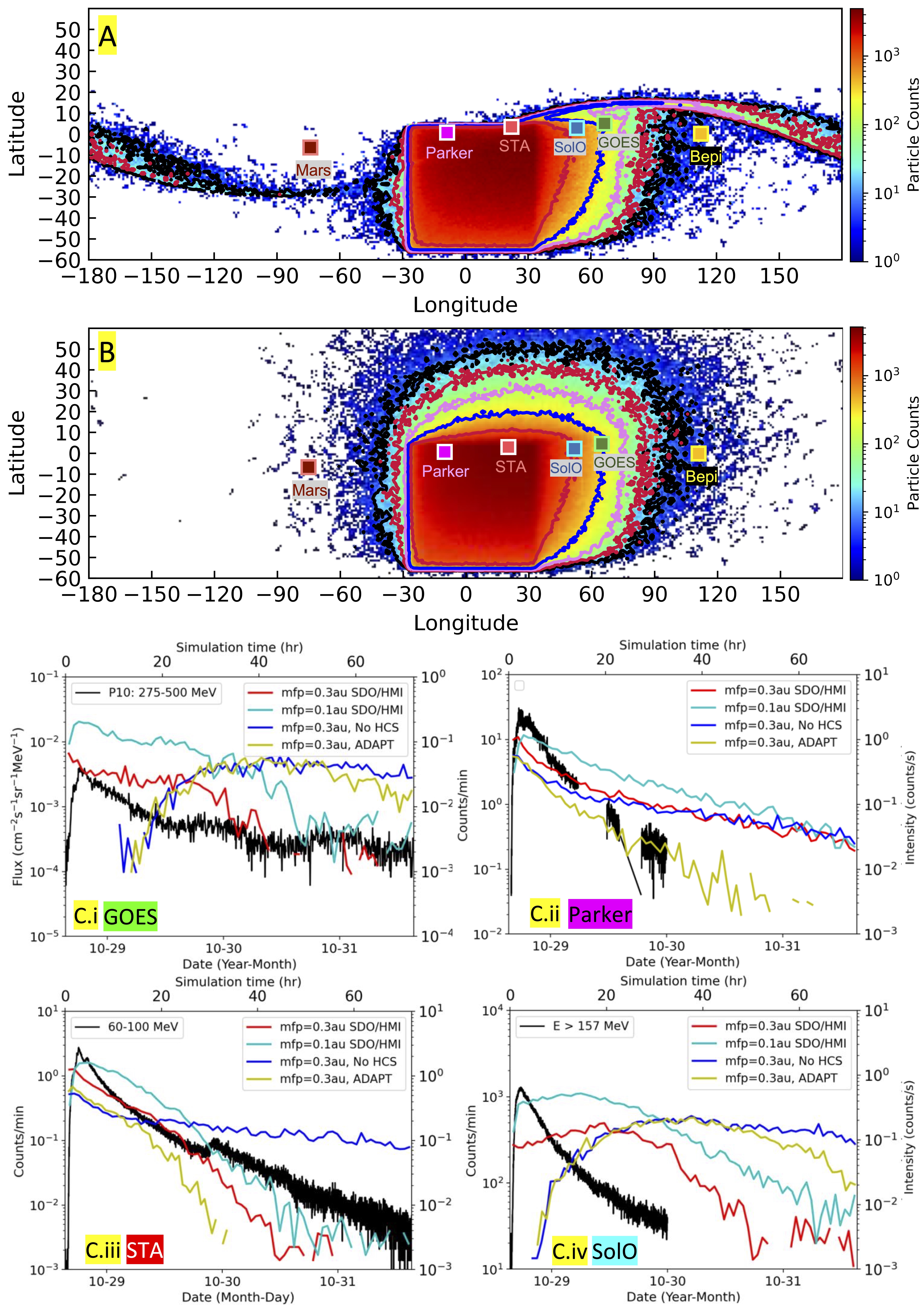}
\caption{Results from the SEP modelling. Top panels (a): 1~au cumulative proton crossing map for 3D test particle simulation of GLE 73. The location of the HCS is from the SDO/HMI model, and the injection region is centred at the flare location, W02 S26. (b): Crossing map with HCS removed. Observer footpoints are as follows: Earth (blue square), SolO (green square), Mars (brown square), Parker (purple square), STA (red square), Bepi (yellow square). 
Bottom panels:
(c) Observations (black) with simulation results for no HCS (blue), 60$\times$60$^{\circ}$ injection for the SDO/HMI HCS and 0.3au mean free path (red), 60$\times$60$^{\circ}$ injection for the SDO/HMI HCS and 0.1au mean free path (cyan), and 60$\times$60$^{\circ}$ injection with 0.3au mean free path for the ADAPT HCS (yellow). Observers shown in Ci-iv are for GOES, Parker Solar Probe, STA, and SolO.} 
\label{fig:SEPmodelling}
\end{figure}

{Fig.~\ref{fig:SEPmodelling} A shows an example of the cumulative proton crossing map over 72 hours at 1~au for the simulation we conducted for this SEP event with a HCS, for protons in the energy range 50--1000 MeV. Figure~\ref{fig:SEPmodelling} B shows the same simulation with the HCS removed. The HCS is modelled with a fit to the ADAPT HCS and SDO/HMI configurations from Fig.~\ref{fig:PFSS_connect}, with Fig.~\ref{fig:SEPmodelling} A representing the SDO/HMI HCS. The protons are instantaneously injected at 2 solar radii over a region of 60$\times$60$^{\circ}$ in longitude and latitude, as shown for example in Fig.~9A, with uniform spatial distribution over the region.  When the HCS is present, the high-energy protons undergo significant drift longitudinally along the HCS in a westward direction. The longitudinal drift is severely reduced when the HCS is removed. The polarity of the heliospheric magnetic field at the time of the event was $A^+$, so that protons starting away from the HCS move towards it due to gradient and curvature drift.}


{The flux profile over 72 hours from injection is obtained at all spacecraft locations and compared to the observations. The results from GOES, Parker Solar Probe, STA, and SolO are shown in Fig.~\ref{fig:SEPmodelling} Ci-iv. Shown here are the simulated flux profiles for the models with the SDO/HMI HCS fit (mean free path 0.3 and 0.1 au), with the ADAPT HCS (mean free path 0.3 au), and without a HCS (mean free path 0.3 au). In some cases there is a delay in the SEP onset of the simulations, reduced when the mean free path is reduced and the SDO/HMI HCS fit is used. For observers closer to the injection region (i.e. those shown in Fig.~\ref{fig:SEPmodelling} C), the flux profiles for either HCS simulation generally have a closer fit to observations than without the HCS. There is a faster decay in the simulation when the HCS is included, compared to the no-HCS case. 
Flux profiles at Mars and Bepi Colombo (not shown) are very noisy due to low statistics in the simulations. No particles reach the locations of  Mars and Bepi  when the HCS is removed from the simulations. When the HCS is included, particles do reach these locations; however, the counts are minimal.}

Within the assumptions of the model used, the results of the simulation suggest that the HCS could play an important role in explaining the observed extent of the high-energy SEPs to the distant magnetically connected observers. The fact that the results have a strong dependence on the HCS model used  highlights the importance of an accurate HCS description and information on the proximity of the observer footpoints to the HCS, which needs to be considered in future test particle simulations. Turbulence-associated  cross-field transport, not included in our simulations, is likely to facilitate particle transport to the HCS and to observers that are less well connected. 
Simulations including this effect will need to be carried out in future.

\section{Discussion}


\subsection{Solar energetic particle observations}

Using SEP observations from instrumentation on board multiple and widely distributed spacecraft in the heliosphere spanning $\sim$360$^\circ$ in heliolongitude and 0.4 to 1.6~au in heliocentric distance, including the two new solar missions (SolO and PSP), we studied the evolution of SEPs close to the Sun and their wide distribution into the heliosphere. The multi-spacecraft SEP observations show a prompt increase for most of the observers. All the spacecraft observed high-energy protons (i.e. E$>$100 MeV) for this event (see Fig.~\ref{fig:sc_pos} \& Fig.~\ref{fig:GOEs_spectra}). In situ particle observations by six widely separated observers (PSP, STA, SolO, GOES, SOHO, and BepiColombo) and MSL/RAD on the surface of Mars and NMs on the face of the Earth provide direct evidence of the wide spread of SEPs during GLE73. Reconstructed fluxes from SolO/HET and SOHO/EPHIN showed that protons up to $\sim$300 MeV and $\sim$600 MeV were clearly measured during the event. Although  Mars was   almost directly opposite to the  Earth, it also  recorded very clear signatures of the arrival of the particles at its surface (see Fig.~(\ref{fig:sc_pos})). 

From the SEP onset times, as well as the  SRTs determined by VDA and TSA (see Table \ref{tab:VDA}), a relation to the different phases of the solar event indicates four findings. First, the first high-energy protons observed at PSP were released with the first group of type III radio bursts observed by Wind/Waves \citep[see Fig.~12 in][]{Klein2022}. The SEP release from the VDA is almost concomitant to the start of the type II radio burst low in the corona, which is suggestive of the high-energy protons observed at PSP being accelerated at the shock wave that formed promptly in the low corona. The type II emission is probably produced in regions where a supercritical and quasi-perpendicular shock wave has formed \citep[see][]{Kouloumvakos2021, Jebaraj2021}. As we discuss later, the results from the connectivity and the shock modelling give further support to the scenario that the observed SEPs at PSP accelerated at the shock wave since strong regions were connected to the observer from the beginning of the magnetic connection. From the 3D reconstruction of the shock wave we find that, around the SEP release time, the shock apex was located at $\sim$1.5~R$_\sun$ and had a speed of $\sim$1400~km/s. Inside the SEP release time window there is the peak of the HXRs; based on the close proximity of the connected field lines to the AR, this makes it difficult to rule out a flare contribution in the SEP acceleration at this stage.
    
Second, at STEREO-A the release of high-energy protons (from VDA) occurs about seven minutes after the peak of the hard X-ray emission (also at the peak of SXRs and later than the release at PSP) during the second group of type III radio bursts. At this time the EUV wave is clearly observed by SDO/AIA to propagate coherently in the low corona (e.g. Fig.~\ref{fig:remote_sensing} and \cite{Hou2022}). The shock continued its fast expansion in the corona. From the 3D reconstruction, we find that when the shock was located at $\sim$2.5~R$_\sun$ the apex had a speed of $\sim$1800~km/s, the flanks a speed of $\sim$1600\,--\,1150~km/s, and below the flanks in the low corona the shock propagates more slowly \citep[see][]{Hou2022}. From the shock modelling strong shock regions were connected to the observer during the SEP release time window, which suggests that the shock could be responsible for the acceleration and release of SEPs at this location (see further discussion in Sect.~\ref{sec:Disc_model}).
    
Third, for SolO we find that the first high-energy protons (from VDA) were released 12 minutes after the peak of the hard X-ray emission (when the impulsive flare emission is close to the background and 5 minutes after the peak of SXRs) during the third and last group of type III radio bursts. From the  TSA of the highest energy channels of SolO we find a release of the mildly relativistic protons at 15:45~UT. Around the time of the SEPs release, the shock apex is at $\sim$4~R$_\sun$ and the speed is close to maximum. From the 3D reconstruction we find that the shock speed at the apex is $\sim$2075~km/s, and the shock modelling shows that shock regions of moderate strength were connected to SolO; therefore, the shock could have accelerated and released the high-energy SEPs at this observer. Parts of intermittent and patchy type II radio emission can also be seen in the decametric range. Composition observations from the SIS instrument shows no serious enhancement of flare accelerated material, suggesting that a  flare-related contribution, if any, may not have an important role.

Finally, at near-Earth space, the release of the high-energy protons from the  VDA of the SOHO SEP data is near the end of the third group of type III radio bursts and is further delayed with respect to the ultra-relativistic particles observed by SOPO NM resulting in GLE73 and from the mildly relativistic measurements observed in space (i.e. SOHO/EPHIN; see Table \ref{tab:VDA}). The TSA shows that the very high-energy particles ($>$430 MeV) that were recorded in the near-Earth space were released no later than $\sim$11 min after the peak of the flare. The shock wave during the SEP release time window is connected to the observer and is still fast; however, the 3D shock model shows that only weak shock regions were connected to near-Earth spacecraft (see Fig.~5 and discussion in Sect.~\ref{sec:Disc_model}). Therefore, the shock may have contributed to the release of the first  high-energy protons to arrive at Earth, but for the local acceleration of high-energy SEPs at the field lines connected to Earth its role is ambiguous. The composition properties at this location are similar to what is observed at SolO. 

The connection of the SEP release times with the different phases of the solar event, and the results of the shock reconstruction and 3D shock model, suggest that the shock wave had an important role in the acceleration and release of the high-energy SEPs for most of the observers (PSP, STEREO-A, SolO) except for Earth. We explored this aspect further by analysing the SEP spectral and composition properties, which can provide valuable information about the acceleration and transport processes involved. The SEP spectrum of many events comprises  two inverse power laws separated at the `break energy' \citep{Kiselev2022}. Spectra of this nature usually have a flatter slope below the break energy and a steeper slope above it, and are commonly observed in a wide range of energies spanning from a few   to several hundred  MeV \citep[e.g.][and references therein]{Tylka2001}.

The double power law of the peak proton flux energy spectrum shown in our study can be attributed to either a shock acceleration \citep[see e.g.][]{Tylka2001, Tylka2005} or to transport effects \citep[][]{LiLee2015}. Additionally, the exponential rollover in the form of an inverse power law \cite{Ellison1985} that is observed in our differential spectrum suggest that a diffusive shock acceleration applies (see Fig.~\ref{fig:GOEs_spectra}). Hence, the shock-related process could have a significant contribution to the proton acceleration. However, this should not rule out a possible additional particle acceleration in the flare.

Particles of different energy ranges may be accelerated efficiently by different mechanisms (one assumed to be shock-related and the other flare-related) with distinct signatures in the low- and high-energy parts. For example, the spectral hardness of the low-energy proton component could suggest that a shock-related contribution probably dominates this part of the spectrum \citep[see also][]{Zhang2022}, whereas the flatter spectrum in the high-energy component could be attributed to a flare acceleration processes \citep[see discussion in][]{Kiselev2022}. In another scenario the high-energy component may come from a re-acceleration of the low-energy protons, which were initially accelerated by the flare,  by the evolving coronal shock \citep{Zhang2022}. This could possibly explain why particles may have been accelerated to relativistic energies even past the flare impulsive phase (well after the peak in HXRs) without excluding a flare contribution in the high-energy component.

{The presence of two different accelerators complicates the identification of the sources of SEPs. Nonetheless, if there was a substantial contribution from a flare-related mechanism in this event, additional observational evidence would have been expected in the SEP composition of certain observers.  However, for the majority of the observers measuring the SEP composition during the event, there is no evidence for an increase in $^3$He, nor did the Fe/O composition in SolO/SIS and ACE (and SOHO/ERNE for high energies) significantly differ from that of past large-gradual solar particle events. Therefore, the observational evidence suggests that a flare-related process may not play such an important role, contrary to what alternative scenarios suggest.}

{In the trap-and-release scenario \citep{Klein2022} the particles that are initially accelerated during the impulsive flare phase are trapped in the evolving flux rope of the CME, and escape later through magnetic reconnection with the open magnetic field lines to the observers (e.g. Earth). In this case magnetic reconnection (e.g. a flare-related process) is involved in both the acceleration and later release of SEPs to open magnetic field lines. Observation of flare accelerated material ($^3$He or Fe/O$\geq$1) in the SEP composition during the event would be anticipated in this event if the trap-and-release scenario had a significant role in this particular SEP event, which was not observed by any of the observers. It seems that the trap-and-release model cannot provide a simple solution to the release of the high-energy protons without contradicting the above observations. However, further modelling of this scenario and novel observations from SolO and PSP may help to address some of the above questions and problems.}

\subsection{Connectivity, shock wave, and SEP modelling} \label{sec:Disc_model}

In this study we used multi-viewpoint remote-sensing observations to reconstruct the shock wave and to carry out a detailed analysis of its position and kinematics in 3D. Additionally, we  modeled the shock wave properties (strength and geometry) in 3D. We also estimated  the magnetic connectivity of the various observers performing a field line tracing to the MAS data and we further supplement this analysis using the potential-field source surface \citep[PFSS;][]{Schatten1969, Wang1992} method and different input magnetograms (ADAPT and HMI) (see Section~\ref{subsec:assesment_connect}). Then we estimated the shock parameters at the connected field lines to each spacecraft and we examined the temporal evolution of the shock parameters. At the inferred SEP release times from the VDA, we determined the shock strength at the connected field lines for each observer. 

The results from the connectivity analysis, the shock 3D modelling, and the shock parameters at the connected field lines to each observer indicate the following considerations. First, PSP was magnetically connected {close to the parent AR. The connected field lines were} close to the heliospheric current sheet. This may have played a significant role in the formation of strong shock regions connected to the spacecraft \citep[see][]{Kouloumvakos2019, Kouloumvakos2021}, which would lead to an efficient acceleration of protons. Recent studies have showed the role of the HCS in the SEP shock acceleration \citep[e.g.][]{Kong2017} and also the the role of  high efficiency in transporting SEPs to the observers \citep[e.g.][]{Waterfall2022}. The results from the 3D shock modelling suggest that strong shock regions were connected to PSP throughout the shock modelling interval ($\overline{M}_{fm}$=8.8$\pm$0.9), and therefore the shock would have an important role in the production of the high-energy SEPs observed at this location. At the time of the inferred release time from the VDA the shock was supercritical. Second, for STEREO-A the connectivity analysis showed that the spacecraft was magnetically connected close to the parent AR. From the shock modelling we find that STEREO-A was also connected to strong shock regions {($\overline{M}_{fm}$=7.6$\pm$0.9)} and at the SEP release time window the shock was supercritical. The results suggest that the shock could be responsible for the acceleration and release of the SEPs at this location. Third, for SolO we find that the connected footpoints were located $\sim$55 degrees  from the AR. The connected shock was of moderate strength {($\overline{M}_{fm}$=5.7$\pm$0.8),} and around the release of SEPs from VDA the shock regions connected to this observer were supercritical. The modelling results also support the idea that the shock could be responsible for the observed high-energy SEPs at this observer. There is, however, a discrepancy between the model and the observations. We find that the modelled shock arrives at the connected field lines two minutes after the calculated SEP release time window. This discrepancy probably arises from the accuracy in the connectivity and the 3D reconstruction. Finally, for Earth, the connected footpoints were located $\sim$68 degrees  from the parent AR. At the connected field lines to the observer, the shock wave is found to be weak. From the 3D modelling, we found that on average the shock is supercritical ($\overline{M}_{fm}$=2.8$\pm$0.5) and the peak shock strength during the modelling interval is $M_{fm}$=3.34$\pm$0.4. However, at the SEP release time window the shock is weak ($M_{fm}$=1.2$\pm$0.2), {capable of accelerating and injecting particles, but the} low shock strength values make it difficult for the shock to have a primary role in the acceleration of the relativistic protons observed at Earth without including some additional physical processes or assumptions. We found similar results from the shock modelling for Bepi and Mars, which were also magnetically connected to weak shock regions.

The results of the 3D shock model show that a strong shock developed in the solar corona during the event. Overall, our findings suggest that the shock had a primary role in the acceleration and release of the high-energy SEPs for PSP, STEREO-A, and SolO. For near-Earth, Bepi, and Mars the connected shock regions are supercritical, which means that an efficient acceleration of SEPs (possibly low-energy) was possible; however, the low strength values may not be able to explain the acceleration and release of the relativistic protons. To explore this issue further and investigate the source of the high-energy protons at the distant observers, we carried out 3D test particle simulations of the event using various parametrizations for the width of the injection region and the mean free path. Then we compared the results of the simulations with high-energy observations from all observers, and tried to conclude which parameters gave qualitatively the best results and what physical insight could provide us. From this SEP modelling, we find that a wider injection region than a point source located at the flare region is needed to explain the observed SEP profiles. Additionally, we find that the HCS configuration during the event likely contributed to the efficient transport of particles from a wide injection source causing the wide spread of high-energy SEPs to the distant magnetically connected observers. Overall, this part of our analysis highlights the importance of the proximity of the observer footpoints to the HCS to the efficient transport of SEPs \citep[see further details in][]{Waterfall2022}. This seems to be an important ingredient to explain the SEP releases to the distant connected observers where the connected shock wave regions were weak. It is also able to explain why the relativistic proton event at Earth was observed to be weakly anisotropic. However, the sensitivity of the flux profiles to the shift of the observer footpoints did not allow us to make an in-depth comparison with the observations with a higher level of parametrization. 


\section{Conclusions}

The analysis of the first multi-spacecraft high-energy SEP event of solar cycle 25, which occurred on 28 October 2021, suggests that large widespread high-energy SEP events, including relativistic GLEs, can be dominated by shock wave acceleration \citep[e.g.][]{Cliver2016}. This study shows that the shock wave had an important role and can be considered   the primary accelerator of the high-energy particles observed during this event. In particular, we find the following:

\begin{itemize}
    \item The 3D shock modelling showed that a strong shock wave was formed in the low corona during the event, in agreement with the observations (e.g. radio). \\

    \item Three of the observers (PSP, STEREO-A, and SolO) were connected to strong shock regions throughout the event. For these observers we show that the shock should have a primary role in releasing and accelerating high-energy SEPs. For other observers (e.g. Earth) with a more distant connectivity, the shock wave was weaker at the connected field lines, contrary to the expectations from the presence of high-energy SEPs at these locations. \\

    \item  From the SEP model we  demonstrated that the HCS could have an important role in the efficient transport of high-energy particles throughout the heliosphere from a wide injection source, presumably the shock. \\

    \item Our study cannot exclude a contribution from a flare-related acceleration or release process (e.g. from the flaring region or the CME); however, composition observations from SolO/SIS (and HET) show no evidence of an impulsive composition of suprathermals during the event, which suggests that a flare-related process may not have had such an important role. Similar composition characteristics for this event were also observed for other spacecraft  (see Cohen et al. 2023). At PSP Cohen et al. show that the Fe/O value is somewhat enhanced ($\sim$0.39), but it is clear that the composition at this spacecraft is not dominated by flare acceleration material.
\end{itemize}

In summary, our findings emphasize the importance of a very fast and wide shock, together with the efficient particle transport, to the widespread characteristics of the multi-spacecraft high-energy SEP event of 28 October 2021. As the next step forward a combination of data-driven shock modelling, coupled with SEP acceleration and transport models, can be very useful in analysing and interpreting the underlying physics for other high-energy SEP events. Focusing on the SEP acceleration and transport when the shock is low in the solar corona, a few minutes or tens of minutes after the start of the eruption, is particularly important since most of the high-energy SEPs are produced at low coronal heights \citep[e.g.][]{Gopalswamy2012, Gopalswamy2013}.


\begin{acknowledgements}
AK acknowledges financial support from NASA's NNN06AA01C (SO-SIS Phase-E and Parker Solar Probe) contract.

AP, acknowledge financial support from the European Space Agency's (ESA's) Archival Research Visitor Programme (2022-2023), under the ``Solar orbiTer solAr eneRgetic parTicle eventS (STARTS) catalogue" project (\url{https://members.noa.gr/atpapaio/starts/}) and further recognizes support from NASA's Living With a Star (LWS) project NNH19ZDA001N-LWS.

SD and CW acknowledge support from the SWIMMR project via NERC grant NE/V002864/1 and from STFC (grant ST/V000934/1).

RV and JRP acknowledge financial support from the European Union’s Horizon 2020 research and innovation programme under grant agreement No.\ 101004159 (SERPENTINE). RV acknowledges the support of the Research Council of Finland (FORESAIL, grants 312357 and 336809). JRP acknowledges the financial support by the Spanish Ministerio de Ciencia, Innovación y Universidades FEDER/MCIU/AEI Projects ESP2017-88436-R and PID2019-104863RB-I00/AEI/10.13039/501100011033.

JG and XL thanks the Strategic Priority Program of the Chinese Academy of Sciences (Grant No. XDB41000000), the National Natural Science Foundation of China (Grant No. 42074222) and the CNSA pre-research Project on Civil Aerospace Technologies (Grant No. D020104).

BH, PK, MH, and RWS thank the German Space Agency, DLR, for its support (grants 50OT2002, 50OC2302, and 50QM1701).

CMSC acknowledges support from NASA grants NNN06AA01C, 80NSSC18K1446, 80NSSC18K0223, and JHU.APL173063

This research has made use of PyThea, an open-source and free Python package to reconstruct the 3D structure of CMEs and shock waves \citep{Kouloumvakos2022PyThea}. For the illustration of the spacecraft location of Fig.~\ref{fig:sc_pos}, we used the Solar-MACH tool\footnote{https://solar-mach.github.io} \citep{Gieseler2022}. We would like to thank the Predictive Science Inc. for making their high-resolution MHD data cubes publicly available, which greatly facilitated our research.

Solar Orbiter is a mission of international cooperation between ESA and NASA, operated by ESA. The Suprathermal Ion Spectrograph (SIS) is a European facility instrument funded by ESA. The SIS instrument was constructed by the JHU/Applied Physics Lab, with assistance from CAU Kiel. Parker Solar Probe was designed, built, and is now operated by the Johns Hopkins Applied Physics Laboratory as part of NASA’s Living with a Star (LWS) program (contract NNN06AA01C). Support from the LWS management and technical team has played a critical role in the success of the Parker Solar Probe mission. The Integrated Science Investigation of the Sun (IS$\sun$IS) Investigation is a multi-institution project led by Princeton University with contributions from Johns Hopkins/APL, Caltech, GSFC, JPL, SwRI, University of New Hampshire, University of Delaware, and University of Arizona.

The STEREO SECCHI data are produced by a consortium of RAL (UK), NRL (USA), LMSAL (USA), GSFC (USA), MPS (Germany), CSL (Belgium), IOTA (France), and IAS (France). SOHO is a mission of international cooperation between ESA and NASA. The SDO/AIA data are provided by the Joint Science Operations Center (JSOC) Science Data Processing (SDP).

\end{acknowledgements}


\bibliographystyle{aa}

\appendix

\section{Very high-energy SEP data from SolO/HET \& SOHO/EPHIN} \label{appendixA}

Here we present the recordings of the high-energy channels from SolO/HET and SOHO/EPHIN included in this study (see Fig.~\ref{fig:GOEs_spectra}).

\begin{figure}[h!]
\centering
\includegraphics[width=0.74\columnwidth]{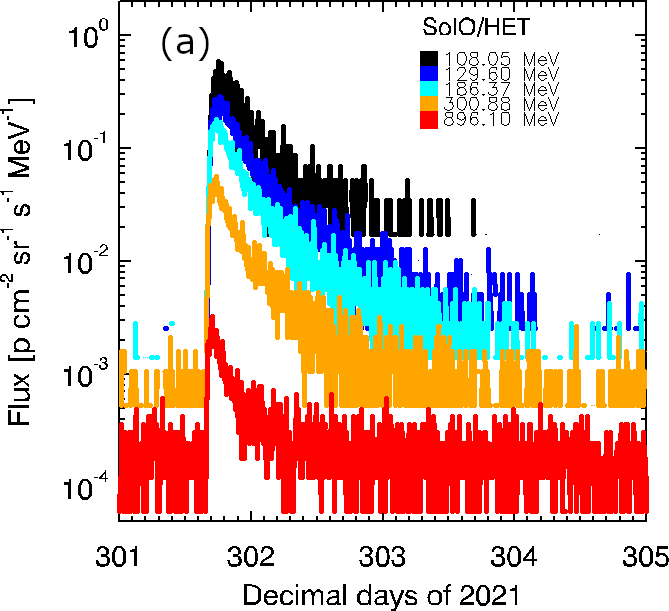}
\includegraphics[width=0.74\columnwidth]{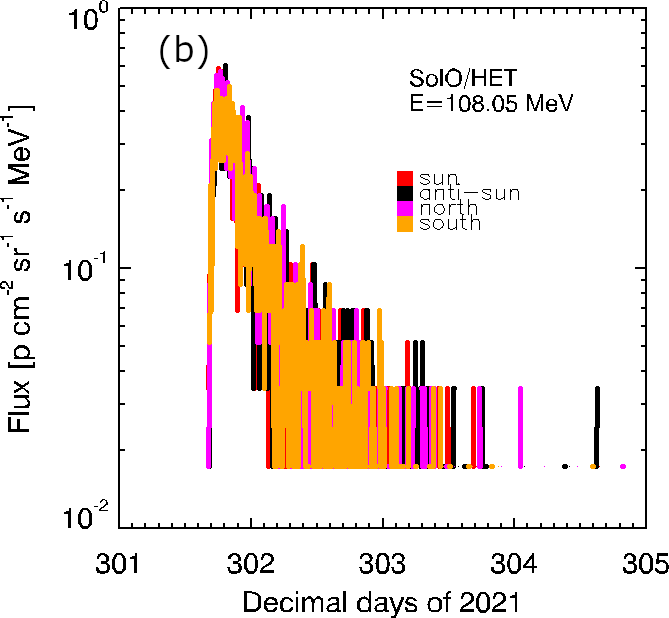}
\includegraphics[width=0.74\columnwidth]{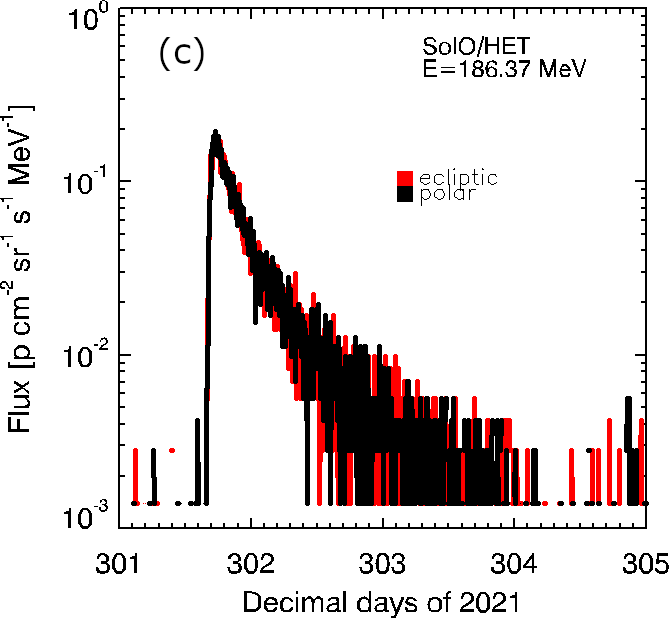}
\caption{ High energy SEP observations from SolO during the event. (a) The reconstructed data are in the range 100-900 MeV and there are five bins indicated by different colours in the plot; (b) Directional information for the first bin of SolO/HET (108.05 MeV). Reconstructed measurements in the sun (red), anti-sun (black), north (magenta), and south (orange) directions are shown; (c) Reconstructed measurements for the third bin of SolO/HET (186.37 MeV) in the ecliptic (red) and the polar (black) directions are shown.} 
\label{a1}
\end{figure}

For SolO/HET there are five bins in total (see Fig.~\ref{a1}(a)). The first two bins that span 100-130 MeV (108.05, 129.60 MeV) are measured at four directions (i.e. sun, anti-sun, north, south) (see an example in Fig.~\ref{a1}(b)). The three highest  bins span  180-900 MeV (186.37, 300.88, 896.10 MeV) and have no directional information. The particles are measured in the ecliptic covering both the sun and anti-sun directions assuming an isotropic flux (within the uncertainty) and in the polar direction (see Fig.~\ref{a1}(c)). In addition, the highest channel (896.10 MeV) has a very complicated and broad response and is sensitive to electrons above 15~MeV. The energy channel at 300.88 MeV presents an onset at $\sim$15:49~UT.

\begin{figure}[h!]
\centering
\includegraphics[width=0.74\columnwidth]{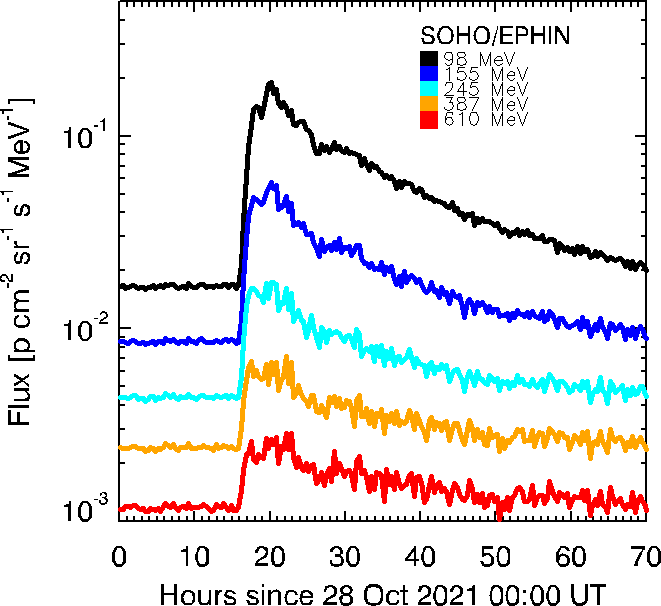}
\caption{GLE73 as  observed in the high-energy bins of SOHO/EPHIN. The reconstructed data span  98-610 MeV and there are five bins indicated by the different colours in the plot.} 
\label{a2}
\end{figure}

For SOHO/EPHIN \cite{Kuhl2015} has shown that very high energies can be identified by the instrument. In particular, Table~2 of \cite{Kuhl2017} shows that the geometric mean energy covered by EPHIN spans from 62 to 610 MeV, with the recordings above this point being contaminated by electrons (and thus disregarded). Figure~\ref{a2} presents these reconstructed fluxes for five mean energies from 98 to 610 MeV for GLE73. The highest energy channel (i.e. 610 MeV) presents an onset at $\sim$15:50~UT.


\section{Estimating the SEP onset and release times} \label{appendixB}

\begin{figure*}
\centering
    \includegraphics[width=\textwidth]{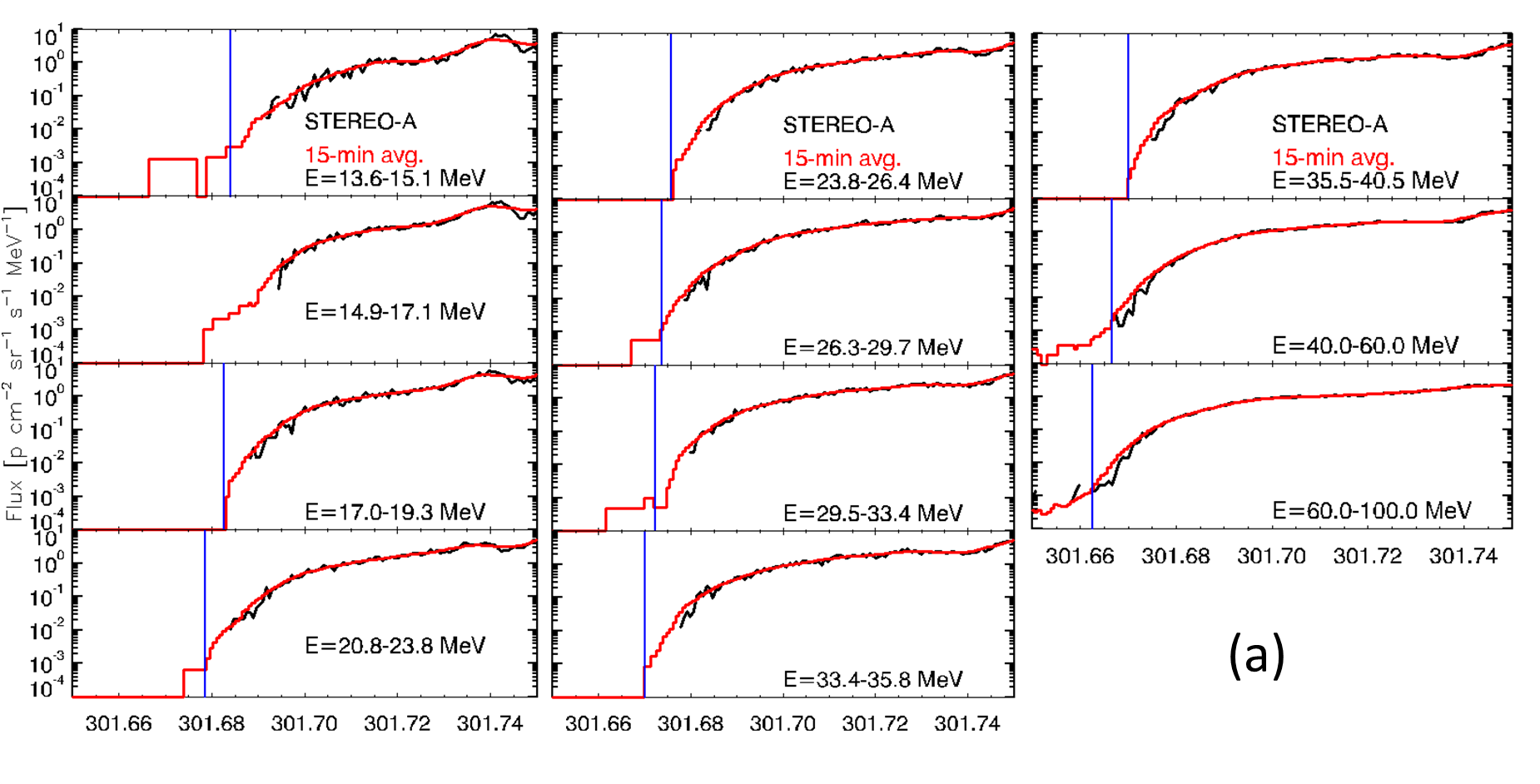}
    \includegraphics[width=\textwidth]{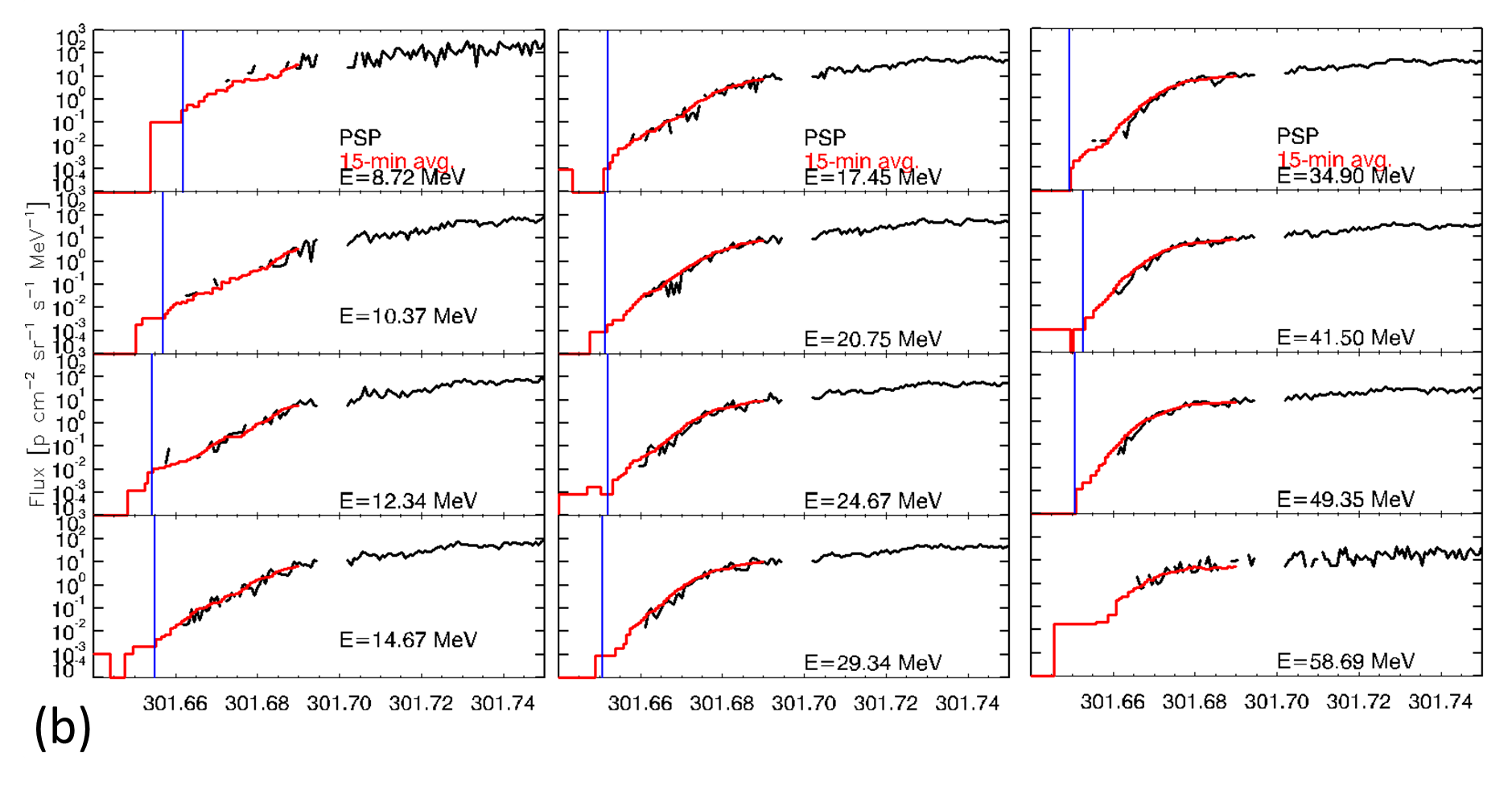}
\caption{Onset of the 28 October 2021 SEP event as observed in the proton energy channels of (a) STA/HET and (b) PSP/HET. The black lines are one-minute averages, whereas the red lines are long-term averages (as indicated in the panels). The blue vertical lines indicate the onset time identified per channel and used in the analysis.} 
\label{fig:onsets}
\end{figure*}

To determine the onset times, we used two different methods when  possible, namely the Poisson-CUSUM method (PCM) or the $n-\sigma$ criterion (SC). The SOHO/ERNE onsets were obtained by the PCM and the  SolO/HET onsets by the SC method that worked the best. For STA both methods resulted in very delayed onset times. Therefore, using the background at high energies (E=60\,--\,100 MeV), we estimated the onset times by a linear fit to the rise phase of the lower-energy channels. For PSP it was not possible to get reliable results (i.e. onsets) with any of the above methods, and  therefore the onset times were identified by eye as follows: longer-term averaged intensities were used and the onset was marked at the time that statistically significant fluxes above the pre-event intensities (by a factor of $\sim2\sigma$) were identified, while the intensities kept rising from this time onwards \citep[see also a similar approach in][]{Lario2017}. The identifications for PSP remain relatively ambiguous, especially for the highest energies (i.e. $E_{mean}$=58.69 MeV). Nonetheless, PSP clearly measures the event up to high energies (see Fig.~\ref{fig:sc_pos}).

To determine the SEP release times we used VDA and TSA. VDA is a method that is based on the determination of the SEP onset times at different energies and that presents these onset times as a function of their inverse velocity (1/v) at the respective energies. To employ the VDA an ordinary least-squares (OLS) linear fit was consequently applied to the identified onset times and the inverse velocities of the channels employed in this work. The underlying assumption of VDA is that all particles, at all energies, are released simultaneously from their source, and thus those particles with higher energies (i.e. speeds) will arrive first at the observer experiencing little scattering. Here VDA was applied to the measurements of PSP/EPI-Hi/HETA, STEREO/HET, SolO/EPD, and SOHO/ERNE. On the other hand, TSA was applied to the reconstructed fluxes of the very high-energy particles from SolO/HET and SOHO/EPHIN, to the fluxes of GOES/SEISS, and to the identified onset time of GLE73 at Mars/RAD measurements. TSA involves shifting the observed proton event onset at each spacecraft and/or channel back to the Sun \citep[see details in][]{Paassilta2018} and provides an upper limit for the release of the high-energy particles \citep[see discussion in][]{Vainio2013}.

{When utilizing VDA and/or TSA, a limitation arises when determining the onset times. This is especially pertinent in the
case of (a) weak SEP events with low statistics, (b) slowly rising SEP events, and (c) if high background fluxes are present in the particular energy channel due to preceding SEP events. All of this makes the coherent identification of a reliable onset time challenging. Additionally, the observed in situ onset time can be affected from the field of view of the observing instrument, in case this is restricted and not closely aligned with a beam of first-arriving particles, and thus from the obtained coverage of the intensity distribution(s) with respect to the magnetic field vector (i.e. pitch-angle) \citep[see also][]{Lario2017}. As can be seen in Figure \ref{fig:onsets}, all onset times were identified as close to the start of the enhancement as possible. For both PSP and STA traditional methods resulted in delayed onset times. This possibly is attributed to the lack of high-resolution data during the start of the event and during the pre-event background period, for each channel. It was not possible to obtain a reliable onset time for the second HET channel ($E$=14.9-17.1 MeV; see Figure \ref{fig:onsets}) of STA and the last channel of PSP ($E_{mean}$=58.69 MeV; see Figure \ref{fig:onsets}), which had a large data gap right at the start of the event, making the identification of the onset time challenging. Thus, these two channels were not taken into account in the VDA.} 

\section{Assessment of the connectivity estimates} \label{subsec:assesment_connect}

{We employed an alternative approach to evaluate and assess the connectivity estimates, distinct from the field line tracing method employed in Section~\ref{sec:shock_model}. In this case we determined the magnetic connectivity between the observers and the solar surface by utilizing the techniques and methods described in \citet{Rouillard2020}. For the interplanetary magnetic field we followed the same procedure as in Section~\ref{sec:shock_model} and assumed a simple Parker spiral, whereas for the low-coronal part we used the PFSS model instead of the field line tracing of the MAS data.} For each observer, using the Parker spiral we calculated the location of the footpoints at the source surface, {which is the outer surface boundary where the magnetic field lines are forced to open. The radius of the source surface is typically assumed to be at 2.5 solar radii in heliocentric coordinates, and we used the same value in our study.} Then from this connection point, we performed a field line tracing to the PFSS coronal magnetic field solutions.   We found the magnetic connectivity of the observers below the source surface assuming an uncertainty of five degrees around the nominal connection point at the source surface;   the tracing was performed around this region.

\begin{figure*}[h!]
\centering
    \includegraphics[width=0.85\textwidth]{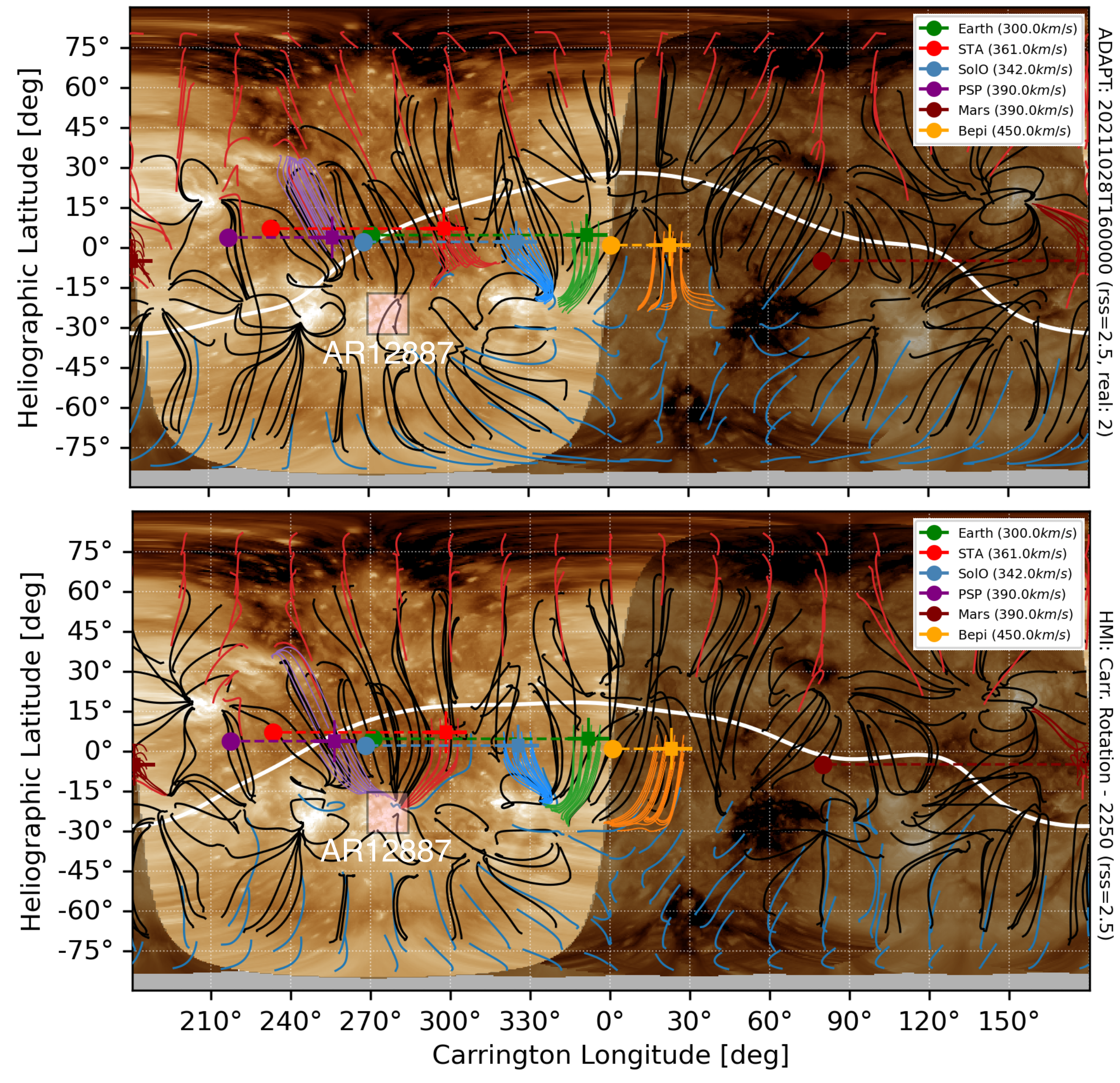}
\caption{ Similar to Fig.~\ref{fig:MAS_connect}, but for the magnetic connectivity of the observers to the solar surface using the PFSS model and two different input magnetograms. The top panel shows the magnetic connectivity based on the ADAPT photospheric magnetic field map on 16:00~UT (second realization) and the bottom panel using the SDO/HMI synoptic maps. The observers' positions are depicted as coloured circles. The Parker spirals are shown with the dashed coloured lines and the footpoints of the spirals at the source surface (2.5~$R_\sun$) are shown with the coloured squares. The traced magnetic field lines from the PFSS model that connect to each observer are depicted with the coloured lines, and the open (red or blue depending on the polarity) and closed (black) field lines from the PFSS model are presented. The neutral line at the source surface is shown with the white line. The background map is constructed using SDO/AIA images at 193~\AA.} 
\label{fig:PFSS_connect}
\end{figure*}

We calculated the PFSS coronal magnetic field solutions for a magnetic map using  pfsspy \citep{Stansby2020}, which is a Python package for PFSS modelling. {We used two different photospheric magnetic field maps as  input to the PFSS model. The first is the Air Force Data Assimilative Photospheric Flux Transport \citep[ADAPT:][]{Arge2010,Arge2013} maps that  are produced using magnetograms from the Global Oscillation Network Group. The second is the SDO Helioseismic and Magnetic Imager (HMI) synoptic maps that are constructed from HMI line-of-sight magnetograms over a full solar rotation. The ADAPT model uses} a flux-transport model \citep{Worden2020} {to simulate} the evolution of the magnetic field for regions where data are not available. This model accounts for time-dependent phenomena such as differential rotation, super-granulation, and meridional flows and calculates the evolution of the magnetic field to provide an updated map for regions on the solar far side. The ADAPT is an ensemble model, so the provided maps consist of 12 different realizations of the solar surface magnetic field. Each realization will give different PFSS solutions, so the location of the heliospheric current sheet and the magnetic connectivity can be slightly different. The best realization can be determined by comparing the model-derived parameters with observations. For example, the polarity of the interplanetary magnetic field measured in situ is compared with the model-derived polarity or by comparing the location of the heliospheric current sheet against WL observations \citep[see e.g.][]{Poirier2021,Badman2022}. We find that the different realizations give qualitatively similar results for the PFSS magnetic field solutions; however, the second realization {of the 12 ADAPT solutions compares qualitatively better} with the WL observations for this event. We use this realization for our connectivity analysis.

{Figure~\ref{fig:PFSS_connect} shows the magnetic field solution from the PFSS model and the results from the magnetic connectivity analysis projected to a Carrington map (see details in Section~\ref{sec:shock_model} and Fig.~\ref{fig:MAS_connect}). For the top panel we use the ADAPT map as input to the PFSS model, while for the bottom panel we use the HMI synoptic maps. In the two maps the positions of the HCSs are similar, with some differences visible in the north-eastern direction from AR12887. The magnetic connectivity of the different spacecraft is depicted in Fig.~\ref{fig:PFSS_connect} with the coloured field lines. The results of this connectivity analysis are similar to the results presented in Section~\ref{sec:shock_model}. More specifically, PSP and STEREO-A have the best connectivity to AR12887, but there are some differences in the connectivity of the two observers depending on the input map used in the PFSS model. From the connectivity analysis based on the ADAPT maps we find that STEREO-A was magnetically connected to a region north-west of AR12887 with CRLN=$317^\circ\pm9^\circ$, whereas PSP was magnetically connected (CRLN$=237^\circ\pm5^\circ$) to a region north of AR12887. On the other hand, using the HMI synoptic maps both PSP and STEREO-A have field lines that were well connected to the source AR12887. For SolO the connected footpoints were around CRLN$=335^\circ\pm4^\circ$ and for Earth around CRLN$=348^\circ\pm5^\circ$. BepiColombo has the most distant connectivity from AR12887, and it was magnetically connected (CRLN$=29^\circ\pm12^\circ$) to the periphery of a small coronal hole that was located at the far side of the Sun. Except for PSP and STA there is no significant difference for the connectivity of any other observer when comparing the results between the two input maps in the PFSS model. Furthermore, the derived polarity from the connectivity estimates are roughly consistent with the in situ polarity of the various spacecraft. Before and around the onset of the event, we find that  PSP and STA polarities are mostly positive, but they bounce between the two polarities probably because their footpoints are close to the HCS. On the other hand, SolO and ACE polarities are mostly negative around the same time, but ACE polarities show some bounces to positive polarity.  }

\begin{figure}[h!]
\centering
\includegraphics[width=0.48\textwidth]{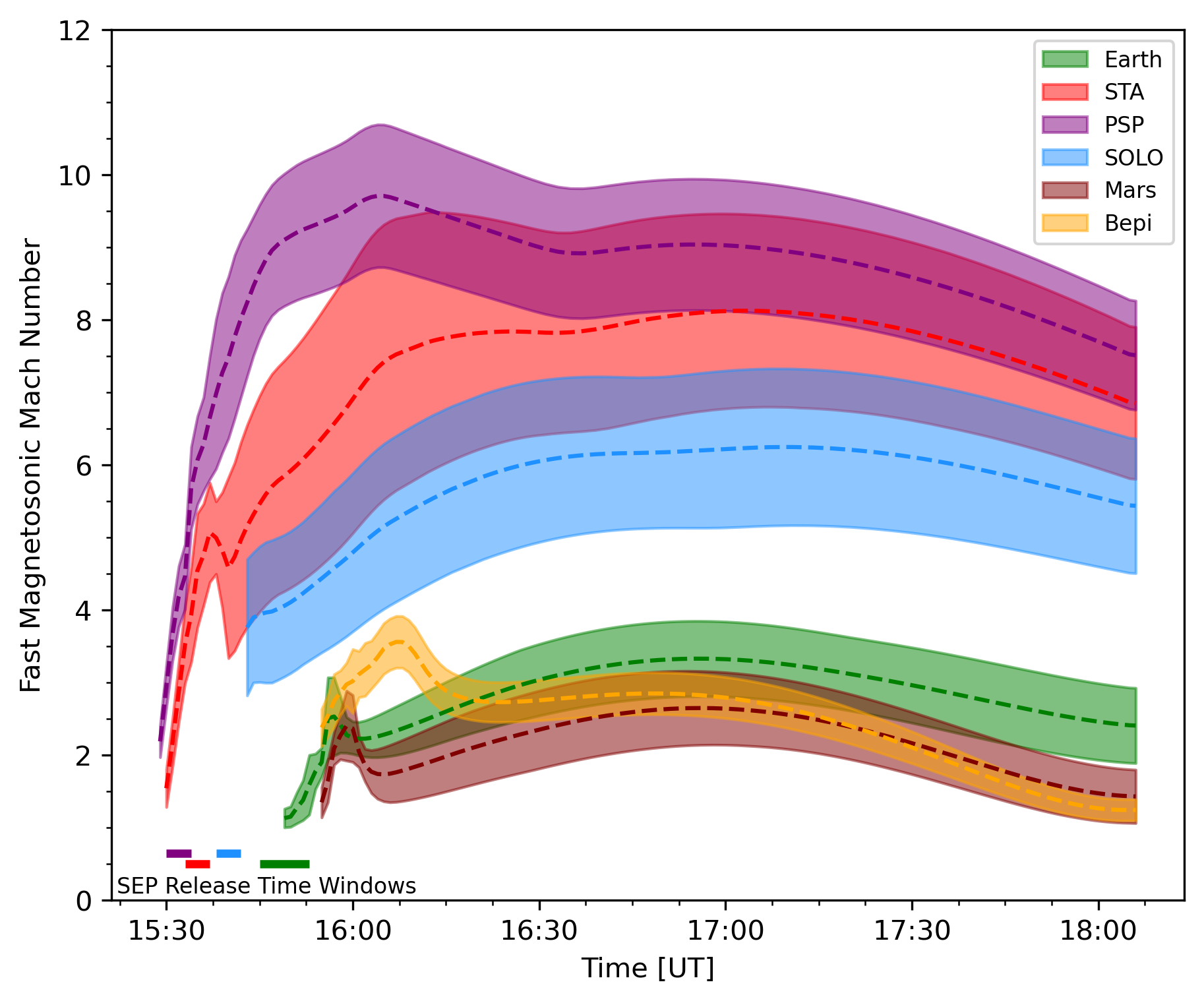}
\caption{ Similar to Figure~\ref{fig:shParam}. The colour-shaded areas depict the variation of $M_{fm}$ using the PFSS connectivity estimates.}
\label{fig:shParam_connect}
\end{figure}

{In Fig.~\ref{fig:shParam_connect}, we show the results for the evolution of the shock $M_{fm}$ using the magnetic connectivity estimates from the PFSS model. We find similar results for most of the observers, except for PSP where the $M_{fm}$ exhibit higher values at the beginning of the modelling and for about 30 minutes. The first connection to PSP is also a few minutes earlier. } Summarizing the above results, we find that all the observers {were at some point connected to a supercritical shock}. PSP, STEREO-A, and SolO were magnetically connected to strong shock regions ($M_{fm}>4$), whereas Earth, Mars, and Bepi were connected to weaker shock regions ($M_{fm}<3$). 

\end{document}